\documentclass[epj]{svjour}
\usepackage{epsfig}
\usepackage[nooneline]{subfigure}
\usepackage{amsmath}
\usepackage{caption}
\usepackage{graphics}
\begin{document}
\title{Random time averaged diffusivities for L\'evy walks}
\subtitle{
}
\author{D. Froemberg \and E. Barkai 
}                     
%
%
\institute{Department of Physics,
Bar Ilan University, Ramat Gan 52900, Israel 
}
\date{Received: date / Revised version: date}
%
\abstract{
We investigate a L\'evy-Walk alternating between velocities $\pm v_0$ with opposite sign. The sojourn time 
probability distribution at large times is a power law lacking its mean or second moment.
The first case corresponds to a ballistic regime where the ensemble averaged mean squared displacement (MSD) 
at large times is $\left\langle x^2 \right\rangle \propto t^2$, the latter to enhanced diffusion with 
$\left\langle x^2 \right\rangle \propto t^\nu$, $1<\nu<2$.
The correlation function and the time averaged MSD are calculated.
In the ballistic case, the deviations of the time averaged MSD from a purely ballistic behavior 
are shown to be distributed according to a Mittag-Leffler density function. 
In the enhanced diffusion regime, the fluctuations of the time averages MSD vanish at large times, yet very slowly.
In both cases we quantify the discrepancy between the time averaged 
and ensemble averaged MSDs. 
} 

\authorrunning{D. Froemberg, E. Barkai}
\titlerunning{Random time averaged diffusivities...}
\maketitle
\section{Introduction}
\label{intro}

Dispersion of Brownian particles is described on the macroscopic level by Fick's second law which states
that the local change in particle density is proportional to the negative gradient of the local particle flux,
where the diffusion constant $D$ is the proportionality factor.
Accordingly, the density of the particles is a spreading 
Gaussian with the mean squared displacement (MSD) going linearly with time, $\langle x^{2}\rangle = 2 D t$. 
Einstein \cite{Einstein05} established a relationship
between the macroscopic quantity $D=\langle (\partial x)^{2}\rangle/{2\langle \tau\rangle}$ and the 
underlying stochastic process with jump lengths 
$\partial x$ of variance $\langle (\partial x)^{2}\rangle$ and the average time passing between two jumps
$\langle \tau\rangle$.
For a single Brownian particle, the time averaged MSD (TAMSD) $\bar\delta^2$ has to be considered,
\begin{eqnarray}
\overline{\delta^2} &=& 
\frac{1}{t-\Delta} \int_0^{t-\Delta} \left[ x(t^\prime + \Delta) - x(t^\prime) \right]^2\; dt^\prime. 
\label{tamsd0}
\end{eqnarray}
 
\noindent
Here the averaging was performed over all displacements along the single trajectory
occuring during a fixed lag time $\Delta$ within the measurement time $t$.
Due to the stationary increments, for a Brownian motion the time averaged MSD
will attain the limit $\overline{\delta^2}=2D\Delta$ for large times $t$.
Brownian motion is therefore ergodic, ensemble and time averages are equal,
$\langle x^2\rangle=\overline{\delta^2}$. It is important to note that in practice 
an estimation of ensemble averaged diffusion constants from single particle trajectories
even for normal diffusion is a subtle issue due to the lack of statistics.
However one can exploit the ergodic property of such processes in order to find 
suitable estimators \cite{osh13-1}, \cite{osh13-2}.

Unlike in normal diffusion, in strongly disordered media
the mean squared displacement often does not grow linearly with time, but according to a power law
$\langle x^{2}\rangle \propto t^{\nu}$.
The case $0<\nu<1$ corresponds to subdiffusion, $1<\nu<2$ indicates enhanced diffusion 
or superdiffusion.
In disordered and complex systems time averages can differ considerably from the 
corresponding ensemble averages. The increasing employment of single particle tracking 
techniques makes it indispensable to understand these differences in order to interpret
the experiments and gain insight to the mechanisms underlying anomalous transport.


Examples for enhanced diffusion are experiments on active transport of microspheres \cite{Caspi02}, 
of polymeric particles \cite{Gal10} or of pigment organelles \cite{Bruno09} in living cells.
Enhanced diffusion in living cells is promoted by molecular motors that move along 
microtubules or the cytoskeleton \cite{Roop04}, \cite{Caspi00}.
In vivo experiments usually examine the trajectories $x(t)$ of single particles and hence assess the
TAMSD Eq. (\ref{tamsd0}) instead of ensemble averages.
Measurements often find this quantity to be a random variable, so that ensemble average
and (single trajectory) time average differ. For this behavior several reasons come into question,
either variations in the probed environments or cells, ergodicity breaking or too short measurement times.
In Refs. \cite{Caspi02}, \cite{Bruno09} the observed 
enhanced diffusion was described within the framework of generalized Langevin equations (GLE), 
which is Gaussian and ergodic \cite{Magd11}. 
Using this theory, the experimentally observed exponents
characterizing the anomalous transport were reproduced. 
However, this Gaussian approach cannot explain the multiscaling of moments found for the enhanced 
motion of polymeric particles in living cells \cite{Gal10}, a feature that
seems to be more consistent with a L\'evy walk scheme.

L\'evy flights describe enhanced diffusion in terms of random walk processes where the distribution of the particle 
displacements lack the second or even the first moment (i.e. give rise to a L\'evy statistics).
L\'evy flights have been used in the past to describe phenomena as diverse as the dispersal of bank notes
\cite{BroGei06}, tracer diffusion in systems of breakable elongated micelles \cite{Ott90} or
animal foraging patterns \cite{Visw96}.
However, L\'evy Flight models can lead to 
unphysical behavior with regard to velocities since L\'evy flights are characterized by extremely large jump lengths,
corresponding to the heavy tails in the jump length distributions.
L\'evy walks address finite velocities by either penalizing long instantaneous jumps with long resting times
(jump models), or by letting the particles move at a certain velocity for a certain time or displacement,
and choosing a new direction (or velocity) and sojourn time according to given probabilities (velocity models) 
\cite{ZumKlaf93}, \cite{West97}.

The theory of L\'evy walks finds a wide range of applications. Experiments with passive tracer particles in a laminar flow
have shown that the flight times and hence displacements within the resultant chaotic trajectories of the tracer particles 
can exhibit power-law distributions \cite{SolSwin93}. Likewise, the motion of tracers in turbulent flows 
can be described as L\'evy walks \cite{ShleWest87}.
Another example is the stochastic description of on- off-times in blinking quantum dots, where 
the intensity corresponds to the velocity of a particle alternating between states of zero and 
constant nonzero velocity. Nonergodic behavior was found in the correlation functions for sojourn time distributions 
lacking their mean \cite{MargBarJCP04},\cite{MarBar05}.
Another example is the dynamics of cold atoms in optical traps \cite{Kess12} 
and the related Brownian motion in shallow logarithmic potentials \cite{Dech12}, or perturbation spreading in 
many-particle systems \cite{Zab11}.
Moreover, also deterministic systems such as certain classes 
of iterated nonlinear maps may show enhanced diffusion.
The chaotic behavior of resistively shunted Josephson junctions 
manifests itself in an anomalous (deterministic) phase diffusion,
which can therefore be modelled by means of such maps \cite{GeiZach85}.
In turn, the enhanced diffusion behavior emerging from such iterated maps 
can be modelled stochastically using the L\'evy walk
approach (in particular velocity models) \cite{ZumKlaf93}.

In this article we study L\'evy walks in one dimension where
the persistence times in the positive- or negative velocity state are drawn according to a 
probability density function $\psi(\tau)$ with either first or second moment lacking. The first case 
is referred to as the ballistic case, the latter as the subballistic or enhanced case.
Recently, the dynamics emerging from nonlinear map similar to the L\'evy walk was investigated \cite{Akimoto12}. 
However, this work did not address the fluctuations of the TAMSD. Fluctuations were considered in a very recent
numerical study for the special case of L\'evy walks exhibiting enhanced diffusion \cite{Ralf13}, and
numerically and analytically for enhanced and ballistic case in a brief publication of the authors \cite{ours}.

The article is divided into four parts. The first one is dedicated to the ballistic case of a L\'evy walk, 
the second one to a L\'evy flight with a step size distribution lacking the second moment,
and the third one to the enhanced L\'evy walk case. 
For both the ballistic and the enhanced case we first review briefly occupation times, propagators 
and ensemble averaged mean squared displacements.
Then we turn to the ensemble averaged quantities such as correlation functions and
ensemble averages of the TAMSDs. Finally, we investigate 
the distributions and properties of the fluctuations of the TAMSDs.
In the second part, we investigate for comparison the L\'evy flight corresponding to the enhanced case.
We also provide the fluctuations of time averages of L\'evy flights, using simple arguments, 
thus adding to the work in \cite{BurWer10} who addressed this problem rigorously and more generally.

\section{Ballistic regime}

We consider a one-dimensional motion of a particle with a two-state-velocity $\pm v_0$ 
where the sojourn times in the states are drawn 
from a probability density function (PDF) $\psi(\tau)$. 
Hence the particle has a velocity $+v_0$ for period $\tau_1$ drawn from $\psi(\tau)$,
after that switches to velocity $-v_0$ and remains in this state for another period $\tau_2$ also drawn from $\psi(\tau)$.
This process is then renewed.
In particular, this PDF is chosen such that it lacks its first moment with a power-law decay at large
times, $\psi(\tau) \sim A/\Gamma(-\alpha)  \tau^{-1-\alpha}$ with $0<\alpha<1$. Particularly in the simulations we will use 

\begin{eqnarray}
\psi(\tau) &=& \left\lbrace \begin{array}{l c c}
\alpha \tau^{-1-\alpha} && \hspace{.5cm} \tau\geq 1 \\
0 && \hspace{.5cm} else\; .
\end{array}\right. .\label{psioft}
\end{eqnarray}

\noindent
The Laplace transform of Eq. (\ref{psioft}) in the small-$u$-limit is

\begin{eqnarray}
\tilde{\psi}(u) &\simeq& 1 - A u^\alpha
\label{psiofu}
\end{eqnarray}

\noindent
with $A=\Gamma(1-\alpha)$ and $u$ being the Laplace conjugate of $\tau$. This relation can easily be derived via
Laplace transformation of $\int_0^\tau \psi(\tau^\prime)d\tau^\prime \simeq 1-\tau^{-\alpha}$, which yields
$\frac{1}{u} \tilde{\psi}(u) \simeq \frac{1}{u}\left(1-\Gamma(1-\alpha)u^\alpha\right)$ 
by using convolution and Tauberian theorems.

In the limit of long times $t$ the particle moves ballistically so that the mean-squared
displacement is $\left\langle x^2 \right\rangle = (1-\alpha)t^2$  \cite{ZumKlaf90}, \cite{Masoliver}.
Recently, a similar system had been generated in the context of deterministic superdiffusion and
the ensemble average of the time averaged mean squared displacement $\langle\overline{\delta^2}\rangle$ was derived \cite{Akimoto12}.
Processes whose temporal dynamics is governed by heavy-tailed PDFs lacking their mean
exhibit ageing \cite{Bou90} which manifests itself in a deviation of the ensemble averaged mean-squared 
displacement from the TAMSDs which are random themselves.
Therefore, the fluctuations of TAMSDs are an important signature of this process.

\subsection{Occupation fraction and particle position}

The distribution for the fraction of time $z_{\pm}=t_\pm/t$ spent in one state (positive or negative velocity) 
after a large time $t$ is given by Lamperti's law \cite{Lamperti}, 
\begin{equation}
p_{occ}(z_{\pm}) = \frac{\sin \pi\alpha}{\pi}
\frac{z_{\pm}^{\alpha-1}\left(1-z_{\pm} \right)^{\alpha-1}}{z_{\pm}^{2\alpha} + \left(1-z_{\pm} \right)^{2\alpha} + 
2\cos\pi\alpha z_{\pm}^{\alpha}\left(1-z_{\pm} \right)^{\alpha}}, \label{lamperti}
\end{equation}
a generalization of the arcsine law which is reproduced for the case $\alpha=1/2$.
The particle position $x(t)$ is given by the integral over the velocities $\int_0^t v(t^\prime) \,dt^\prime$, 
and the temporal mean of the velocity is $x/t=v_0(t_+ - t_-)/t$. 
The distribution of this quantity in the limit of large times had been calculated 
in the context of the mean magnetization of a two-state system \cite{GodrLuck01}. A related problem
is the calculation of the integral over the intensity of emitted light in blinking quantum dots. In that case one state 
corresponds to the ``on''-state, i.e. the emitting state, and the other state to the ``off''-state where 
no light is emitted \cite{MargBarJCP04}.

\begin{figure}[h]
\centering{
{\includegraphics[width=.47\linewidth]{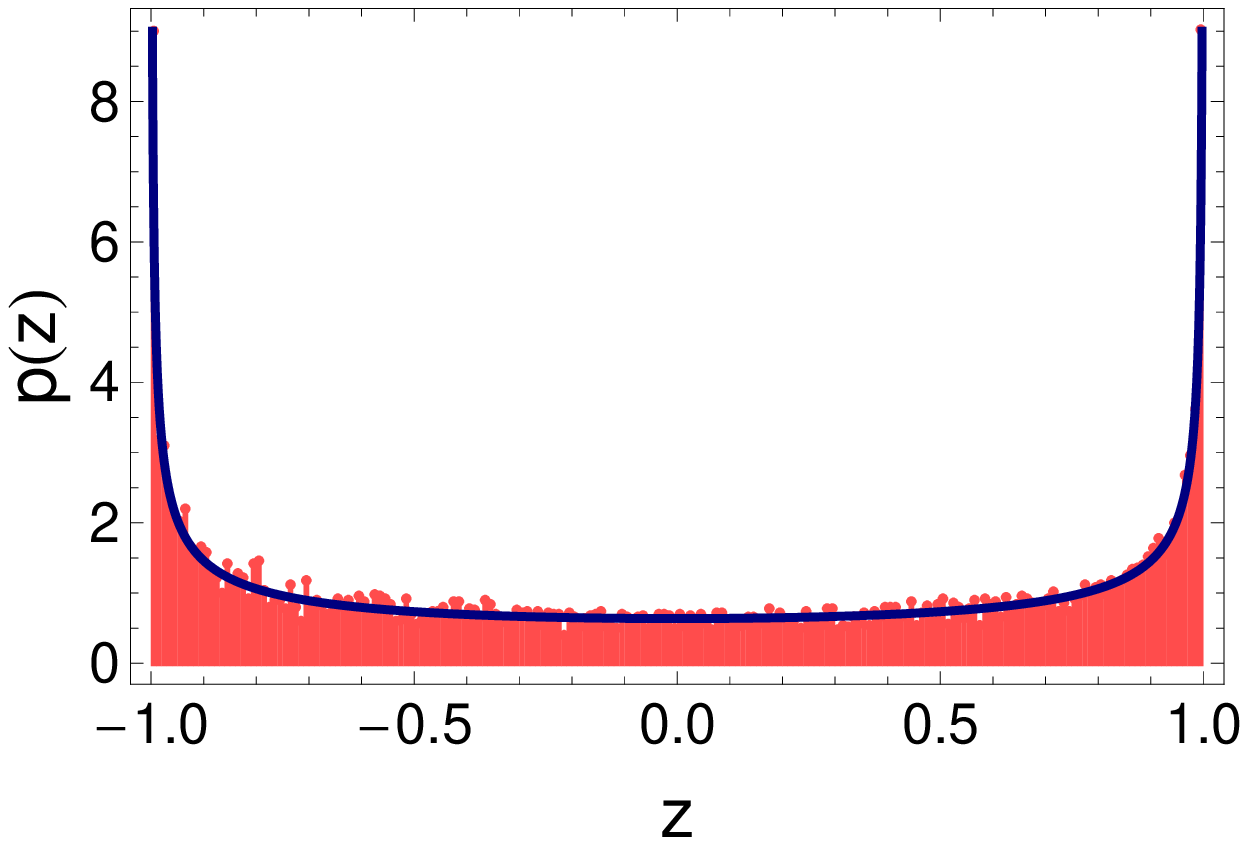}}
\hspace{.05cm}
{\includegraphics[width=.47\linewidth]{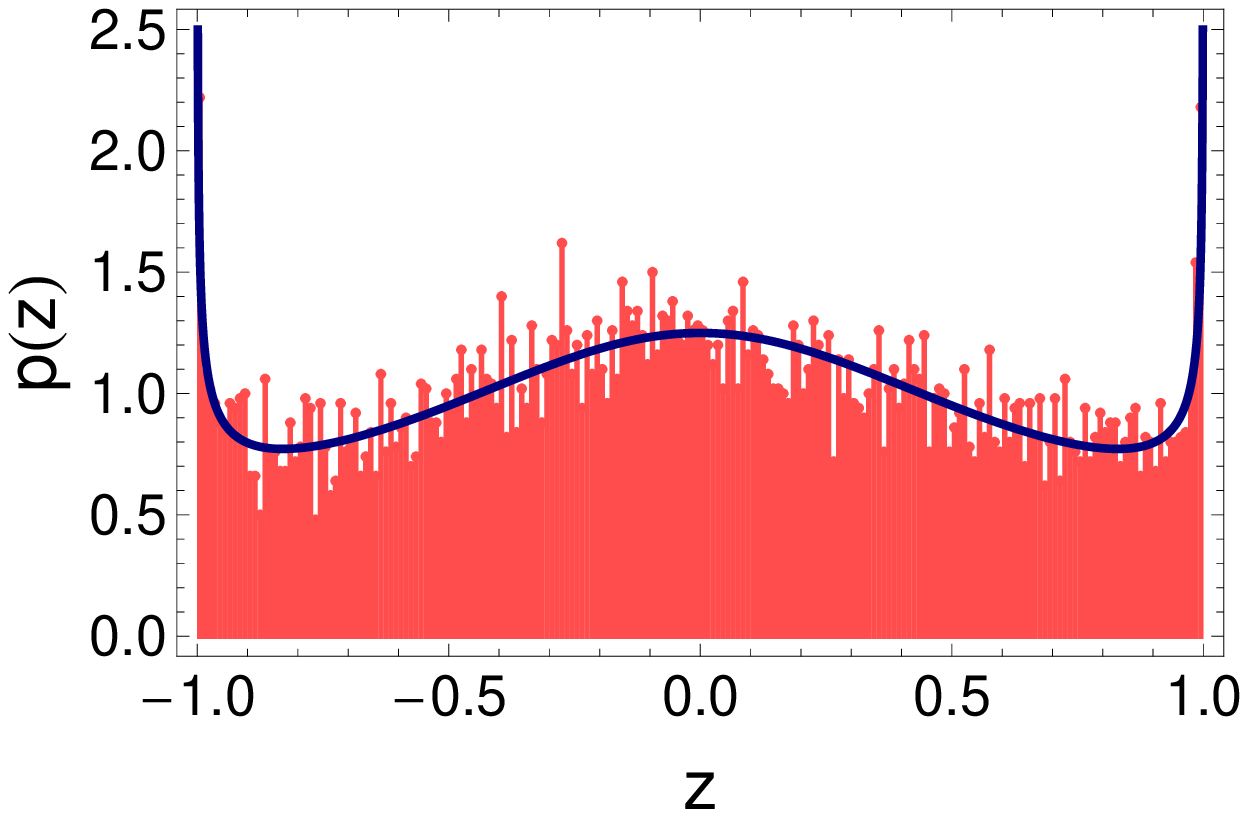}}}
\caption{\label{xdist} Distribution of the position of particles at $t=10^{7}$ for $\alpha = 0.5$ (left) 
and $\alpha = 0.7$ (right panel) in terms of the scaling variable $z$ ($v_0=1$). 
Sample size $10^4$, $\psi(\tau)$ is given by Eq. (\ref{psioft}).}
\end{figure}

The probability to find the particle at position $x$ at time $t$ for large times can be obtained
in terms of the scaling variable $z = x/(v_0 t)$ by using Eq. (\ref{lamperti}) and change of variables.
We have $\frac{x}{v_0 t} = 2 z_+ -1$, hence 
$\frac{\partial z_+}{\partial z} = \frac{1}{2}$ and
$p(z) =\left| \frac{\partial z_+}{\partial z} \right| p_{occ}(z_+)$, 
which finally results in
\begin{equation}
p(z) =
\frac{2\sin \pi\alpha\left(1-z^{2} \right)^{\alpha-1}}{\pi\left(\left(1+z \right)^{2\alpha} + 
\left(1-z \right)^{2\alpha} + 2\cos \pi\alpha\left(1-z^{2} \right)^{\alpha}\right)} .\label{BalProp}
\end{equation}
Fig. \ref{xdist} shows this distribution of the scaled particle position for two different values of $\alpha$.

\subsection{Ensemble average of $\overline{\delta^2}$}

First we will analyze the ensemble averaged TAMSD:
\begin{equation}
\langle \overline{\delta^2}\rangle  = 
\frac{1}{t-\Delta} \left\langle\int_0^{t-\Delta} 
\left[ x(t^\prime+\Delta) - x(t^\prime)\right]^2 dt^\prime
\right\rangle  \; .
\label{TAMSD1}
\end{equation}
Changing the order of integration and ensemble averaging in Eq. (\ref{TAMSD1}), we get
\begin{equation}
\langle \overline{\delta^2}\rangle  =
\int_0^{t-\Delta}  \frac{ \langle x^2(t^\prime+\Delta)\rangle +
\langle x^2(t^\prime)\rangle - 2\langle x(t^\prime)(t^\prime+ \Delta)\rangle}{t-\Delta}dt^\prime.
\label{TAMSD2}
\end{equation}
In order to find $\langle \overline{\delta^2}\rangle$,
we first derive the L\'evy walk correlation function $\left\langle x(t_1)x(t_2)\right\rangle$, 
which turns out to exhibit ageing.
The position correlation function $\langle x(t_1) x(t_2)\rangle $ is related to the 
velocity correlation function $\langle v(s_1) v(s_2)\rangle$ via
\begin{eqnarray}
\langle x(t_1) x(t_2)\rangle 
&=& \left\langle \int_0^{t_1} v(s_1) \,ds_1 \int_0^{t_2} v(s_2)\,ds_2\right\rangle \nonumber\\ 
&=& \int_0^{t_1} ds_1 \int_{s_1}^{t_2} \langle v(s_1) v(s_2)\rangle\,ds_2 \nonumber\\
&& + \int_0^{t_1} ds_2 \int_0^{s_2} \langle v(s_1) v(s_2)\rangle \, ds_1 \label{Cxx0}
\end{eqnarray}
where we took into account that $t_2 > t_1$.
Using the approach of Ref. \cite{GodrLuck01} we obtain for the velocity correlation function
$\langle v(t_1) v(t_2)\rangle$
\begin{equation}
\langle v(t_1)v(t_2) \rangle
= v_0^2 \sum_{n=0}^\infty (-1)^n p_n(t_1,t_2) \label{genCvv}
\end{equation}
where $p_n(t_1,t_2)$ is the probability of the velocity to switch its sign $n$ times within the 
time interval $\left[ t_1, t_2\right]$, i.e. for even $n$ between $t_1$ and $t_2$ we have $v(t_1)v(t_2)=v_0^2$, 
and for odd $n$, $v(t_1)v(t_2)=-v_0^2$.

\begin{figure}[h]
\centering{
{\includegraphics[width=.8\linewidth]{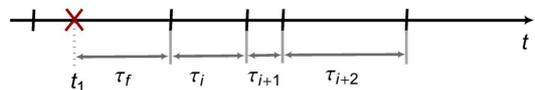}}
}
\caption{\label{fwrec} Sketch of the forward recurrence time $\tau_f$. Ticks symbolize renewal events.}
\end{figure}

In the scaling limit where $t_1$ and $t_2$ are large and Eq. (\ref{psiofu}) applies,
the particle gets stuck in the $+$ or $-$ state for times of the order of the measurement time
due to the lacking first moment of the sojourn time distribution $\psi(\tau)$. Therefore, only the first term 
$n=0$ is relevant in Eq. (\ref{genCvv}).
The corresponding probability $p_0(t_1,t_2)$ of the velocity not switching its sign from a given $t_1$ on up to $t_2$ is
called the persistence probability, to which the velocity correlation function is proportional, 
$\langle v(t_1)v(t_2) \rangle = v_0^2 p_0(t_1,t_2)$.
Let us denote the first waiting time from an arbitrary time $t_1$ up to the next switching event by $\tau_f$,
see Fig. \ref{fwrec}. This first waiting time is called the forward recurrence time, and  
its PDF $\psi_{f,t_1}(\tau_f)$ differs from $\psi(\tau)$ 
since $t_1$ does not necessarily coincide with a renewal event.

\begin{figure}[h]
\centering{
{\includegraphics[width=.44\textwidth]{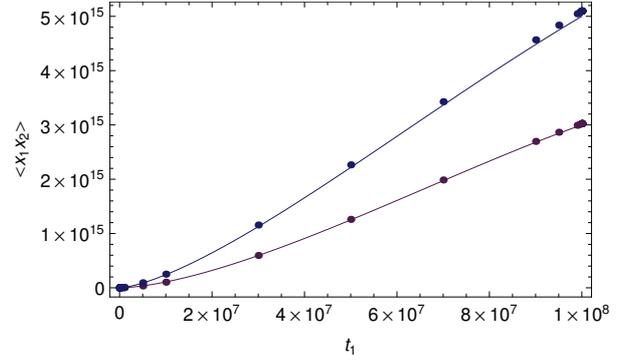}}
\caption{\label{Cxxb} Simulational results for $\langle x(t_1)x(t_2)\rangle$ for $\alpha = 0.5$ (upper) 
and $\alpha=0.7$ (lower graph), 
and the respective theoretical predictions (solid lines), Eq.(\ref{Cxx}). 
$t_2$ was fixed at $10^8$, $v_0=1$. Sample size $10^4$.}
}
\end{figure}

With $t_2=t_1+\Delta$, we express the persistence probability as \cite{GodrLuck01}
\begin{equation}
p_0(t_1,t_1+\Delta) = 
\int_\Delta^\infty \psi_{f,t_1}(\tau_f)\, d\tau_f . \label{genCvva}
\end{equation}
In terms of the scaling variable $\theta=\Delta/t_1$, the pdf of the  forward recurrence time to take a value of $\Delta$
at time $t_1$ reads in the scaling limit where also $\Delta$ is large:
\begin{equation}
\lim_{t_1\to\infty}\psi_{f} (\theta) = \frac{\sin \pi\alpha}{\pi}\frac{1}{\theta^\alpha(1+\theta)} .\label{Dynkin}
\end{equation}
The limit theorem for forward recurrence times $\psi_{f,t_1}$ Eq. (\ref{Dynkin}) is due to Dynkin \cite{Feller}. 
Hence,
\begin{eqnarray}
\langle v(t_1) v(t_1+\Delta)\rangle &=& v_0^2 \int_{\frac{\Delta}{t_1}}^\infty \frac{\sin \pi\alpha}{\pi}\frac{1}{\theta^\alpha(1+\theta)} 
d\theta \nonumber \\
&=& v_0^2 \int_0^{\frac{t_1}{t2}} \frac{\sin \pi\alpha}{\pi}\frac{1}{\xi^{1-\alpha}(1-\xi)^\alpha}\, d\xi \nonumber\\
&=& v_0^2 \frac{\sin \pi\alpha}{\pi} B\left(\frac{t_1}{t_2};\alpha,1-\alpha \right) \label{Cvv}
\end{eqnarray}
where $B(y;a,b)=\int_0^y du {u^{a-1}(1-u)^{b-1}}$ denotes the incomplete Beta-function \cite{AbSteg}. 
Note that this expression yields only real values for $t_2 \geq t_1$.
In the case of $t_1>t_2$, the $t_1$ and $t_2$ in Eq. (\ref{Cvv}) have to be interchanged.
Inserting Eq.(\ref{Cvv}) into Eq.(\ref{Cxx0}) and using integration by parts we find
\begin{eqnarray}
\langle x(t_1) x(t_2) \rangle &=& v_0^2 \frac{\sin \pi\alpha}{\pi} \times 
\left[ t_1 t_2 B\left( \frac{t_1}{t_2};\alpha,1-\alpha\right) \right. \nonumber \\
&&-\frac{1}{2} t_2^2 B\left( \frac{t_1}{t_2};1+\alpha,1-\alpha \right) \nonumber \\
&&\left.-\frac{1}{2} t_1^2 B\left(\frac{t_1}{t_2};-1+\alpha,1-\alpha \right) \right] \nonumber \\
&& - \alpha\frac{v_0^2}{2} t_1^2  \label{Cxx}
\end{eqnarray}
In particular, for $t_1=t_2$ we find the MSD
\begin{equation}
\langle x^2(t_1) \rangle = (1 - \alpha) v_0^2  t_1^2 \, ,\label{msd}
\end{equation}
in agreement with \cite{ZumKlaf90}, \cite{Masoliver}, \cite{GodrLuck01}.
The theoretical autocorrelation functions 
increase in this case with increasing time difference.
For normal diffusion we have $\langle x(t_1)x(t_2)\rangle = 2D \mathrm{min}(t_1,t_2)$, so that 
$\langle x(t_1)x(t_2) \rangle/\langle x^2(t_1)\rangle  = 1$ for $t_2\geq t_1$.
In contrast, the behavior of the L\'evy walk is governed by long periods of ballistic motion. 
Thus, it exhibits strong correlations compared to normal diffusion which are due to the 
long sticking times in the positive or negative velocity states.
In the limiting case $\alpha \to 0$ the particle remains in state $+v_0$ or $-v_0$ throughout 
the measurement so that $x(t) = \pm v_0 t$ with probability $1/2$ for either sign. 
Therefore we expect the purely ballistic, deterministic behavior $\langle x(t_1)x(t_2)\rangle = v_0^2 t_1t_2$
for the position-position correlation function, and hence 
$\langle x(t_1)x(t_2) \rangle/\langle x^2(t_1) \rangle  = t_2/t_1$.
To see this, note that for $\alpha\to 0$, $ B\left( \frac{t_1}{t_2};\alpha,1-\alpha\right)$ 
diverges and the first term in Eq. (\ref{Cxx}) is the only term that remains:
\begin{eqnarray}
&&\frac{ v_0^2 t_1t_2\sin \pi\alpha}{\pi} B\left( \frac{t_1}{t_2}; \alpha, 1-\alpha\right) = \nonumber \\
&&\frac{v_0^2 t_1t_2 \sin \pi\alpha}{\pi} \left(\frac{t_1}{t_2}\right)^{\alpha} \sum_{i=0}^\infty \frac{\Gamma(\alpha+i)}{\Gamma(\alpha)}
\frac{1}{(\alpha+i)i!}\left(\frac{t_1}{t_2}\right)^{i}, \nonumber
\end{eqnarray}
recalling that $t_1/t_2<1$ and taking the limit $\alpha\to 0$ we are left with
\begin{eqnarray}
lim_{\alpha\to 0} \left\langle x(t_1)x(t_2) \right\rangle  = v_0^2 t_1 t_2 \nonumber
\end{eqnarray}
where we used de l'H\^{o}spital's rule. 
Simulations of the system for moderate $\alpha$ show a good agreement with theory Eq. (\ref{Cxx}), 
see Fig. \ref{Cxxb}. 

The asymptotic behavior of the position-autocorrelation function Eq. (\ref{Cxx}) is
\begin{eqnarray}
\langle x(t_1) x(t_1 + \Delta) \rangle &=& \left\lbrace 
\begin{array}{l }
\frac{v_0^2\sin \pi\alpha}{\pi\alpha(1-\alpha^2)} t_1^{2} \left( 1 + \frac{\Delta}{t_1} \right)^{1-\alpha},  t_1\ll \Delta \\
\\
v_0^2(1-\alpha) \left(t_1^2 +\Delta t_1 \right), \hspace{.2cm} t_1\gg \Delta \;.
\end{array}
\right. \label{Cxxapp}
\end{eqnarray}
Note that we made again the transformation from $(t_1,t_2)$ to $t_1,\Delta=t_2-t_1$ and $t_1\leq t_2$.
Inserting the above results for the correlation function Eq. (\ref{Cxx}) and mean squared displacement 
Eq. (\ref{msd}) into Eq. (\ref{TAMSD2}), 
integrating by parts and using again the integral definition of the incomplete Beta function,
we obtain the ensemble averaged TAMSD.
In the limit $\Delta/t \ll 1$ we get
\begin{eqnarray}
\langle \overline{\delta^2} \rangle \approx v_0^2\left[\Delta^2 - 
\frac{\sin \pi\alpha}{\pi\alpha}
\frac{2\Delta^2\left( \frac{\Delta}{t}\right)^{1-\alpha}}{6 - 11\alpha + 6\alpha^2 - \alpha^3}\right].
\label{EAtamsd}
\end{eqnarray}
Note that in fact the short time behavior of the correlation function is not negligible in the integral Eq. (\ref{TAMSD2}) 
since it affects the
long-time behavior of the ensemble-averaged TAMSD $\langle \overline{\delta^2} \rangle$.
It is important to point out that even the first term of $\langle \overline{\delta^2}(\Delta) \rangle$,
Eq. (\ref{EAtamsd}), differs from $\langle x^2(t) \rangle = v_0^2(1-\alpha)t^2$ by a factor. 
This term was also found recently in a different context of deterministic 
maps by \cite{Akimoto12}.

\subsection{Fluctuations of the time averages}
\begin{figure}
\centering{
{\includegraphics[width=.45\textwidth]{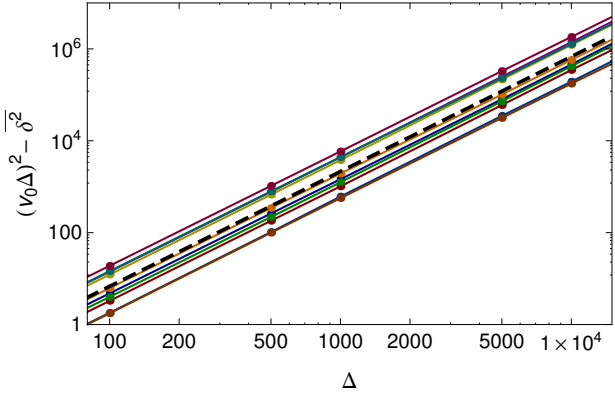}}
}
\caption{\label{fluc} Deviations from ballistic motion of TAMSDs $v_0^2\Delta^2- \overline{\delta^2}$ 
versus $\Delta$ for ten different trajectories; $\alpha=0.5$, $v_0=1$, $t=10^8$. 
Points belonging to the same trajectory are connected by a straight line.}
\end{figure}

More important are the fluctuations of the TAMSDs,
since these allow conclusions to be drawn with respect to the ergodic properties of the system.
Our simulations revealed that these fluctuations are quite small compared to the value of the TAMSDs,
and become even smaller with time relative to the ballistic contribution $(v_0\Delta)^2$.
The fluctuations are more pronounced if one looks at the
shifted TAMSD $v_0^2\Delta^2- \bar{\delta^2}$, which is the
natural random variable of this process as will turn out soon.
In Fig. \ref{fluc} we plot $v_0^2\Delta^2- \bar{\delta^2}$ versus the lag time $\Delta$ for ten different trajectories. 
$v_0^2\Delta^2- \overline{\delta^2}$ remains visibly random. 

\begin{figure}[h]
\centering{
{\includegraphics[width=.45\linewidth]{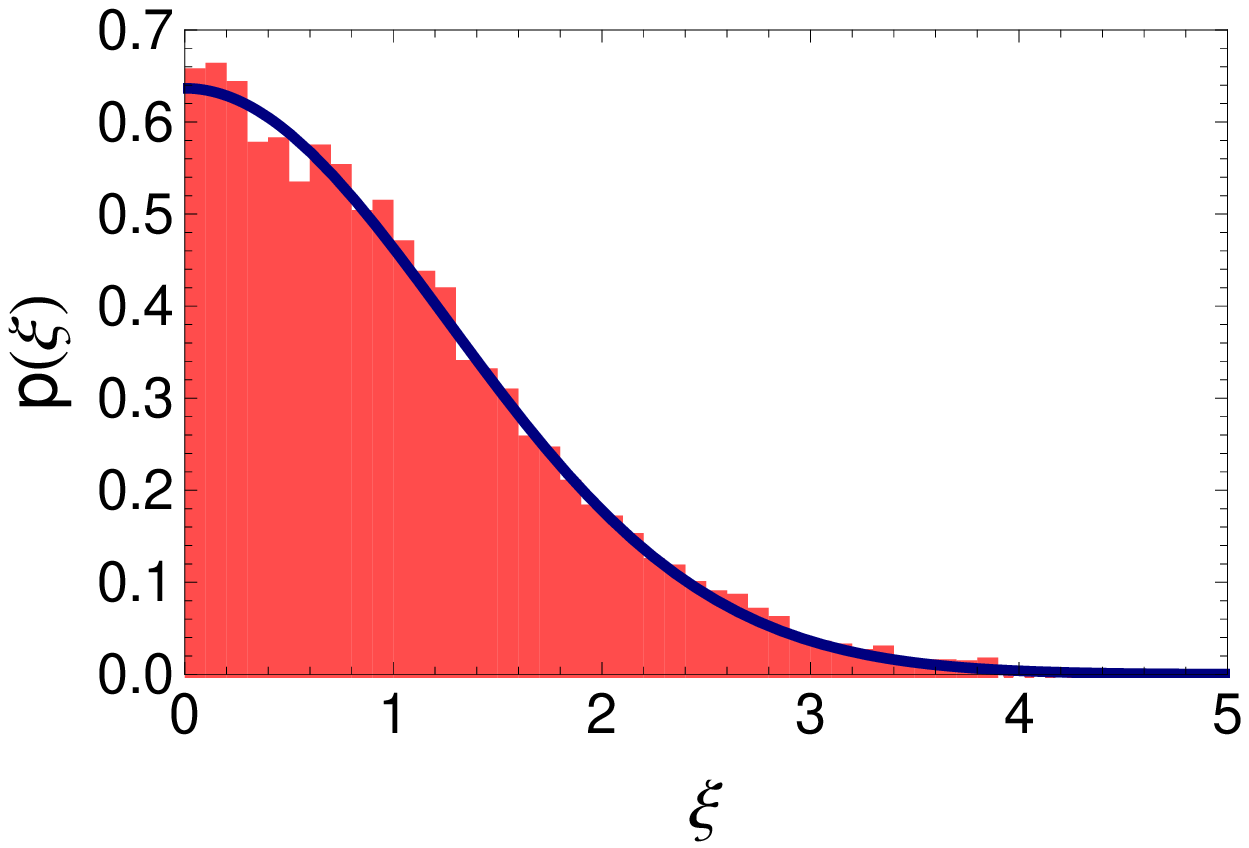}}
\hspace{.1cm}
{\includegraphics[width=.45\linewidth]{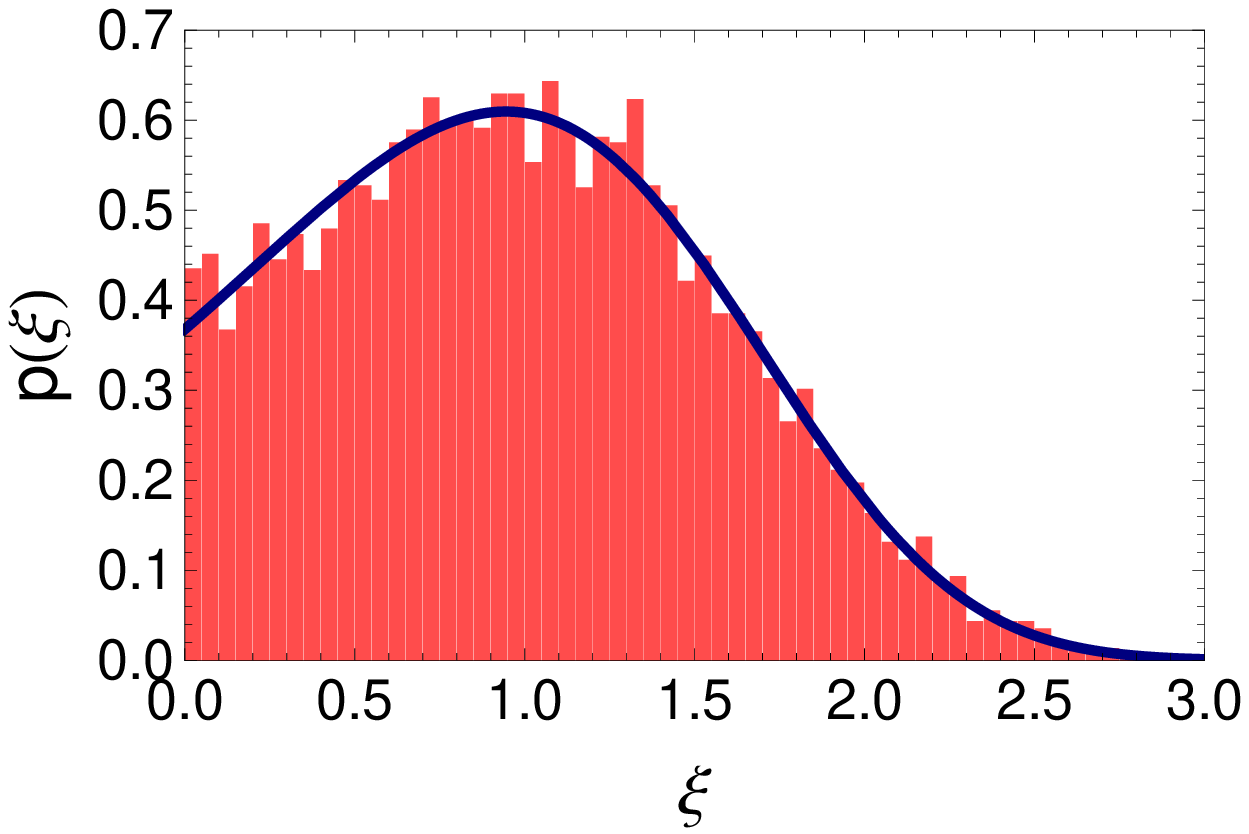}}}
\caption{\label{ML-figa} PDF of 
$\xi = ((v_0 \Delta)^2 - \overline{\delta^2}) / ((v_0 \Delta)^2 - \langle \overline{\delta^2} \rangle) $ 
at $t=10^{6}$ for $\alpha = 0.5$ (left) 
and $\alpha = 0.7$ (right panel); $\Delta=10^3$. 
The histogram shows the result of the simulations, with $\psi(\tau)$ Eq. (\ref{psioft}), 
the solid blue curve the theory Eq. (\ref{ML-th}).
Sample size $10^4$, $v_0=1$.}
\end{figure}

\paragraph{Small $\Delta$ limit:}
In order to specify the distribution of $v_0^2\Delta^2- \overline{\delta^2}$, we construct the following special case.
Consider again the sojourn time PDF Eq. (\ref{psioft}) with $0<\alpha<1$.
Moreover, let us for now only adhere to small $\Delta<1$, 
so that at maximum one change of direction takes place during the interval $\Delta$. 
In this case, the TAMSD Eq. (\ref{TAMSD1}) can be explicitly calculated. 
Clearly, if there were no changes of direction, the TAMSD would be $\bar{\delta^2}=v_0^2 \Delta^2$.
For a single switching event at time $t_s$ within a time interval $(t,t+\Delta)$,
the integrand in Eq. (\ref{TAMSD1}) becomes
\begin{eqnarray}
&&\left[ x(t+\Delta) - x(t) \right]^2 = \nonumber \\
&& \left\lbrace 
\begin{array}{l }
v_0^2 \Delta^2  \hspace{1cm}\text{for}\hspace{.3cm}  t \leq (t_s - \Delta) 
\hspace{.2cm} \text{and} \hspace{.2cm} t_s \leq t \\
(2 v_0 t_s - v_0 \Delta - 2 v_0 t)^2 \hspace{.3cm}\text{for}\hspace{.4cm}  (t_s - \Delta) \leq t \leq t_s .
\end{array}
\right.
 \label{gap}
\end{eqnarray}
Hence, using Eq. (\ref{tamsd0}), one change of direction
within the observation time $t$ reduces the TAMSD to a value of 
$\overline{\delta^2}=v_0^2 \Delta^2-\frac{2}{3} \frac{v_0^2}{t} \Delta^3$ 
for $t\gg\Delta$ ($t-\Delta \approx t$), as shows integration of Eq. (\ref{gap}). Two changes of direction
result in $\overline{\delta^2}=v_0^2 \Delta^2-2\cdot\frac{2}{3}\frac{v_0^2}{t} \Delta^3$ and so forth.
Altogether we find for an amount $n_t$ of switching events within the observation time $t$
\begin{eqnarray}
\overline{\delta^2} &\simeq& \frac{v_0^2}{t}\left[ \Delta^2 t - \frac 2 3 \Delta^3 n_t\right], \label{tamsd-pr}
\end{eqnarray}
where the amount of direction changes $n_t$ within $(0,t)$ is a random variable. 
The probability $p_n(t)$ of the number of events $n$ within $(0,t)$ is determined by the convolution of $n$ 
sojourn time PDFs with the probability of no event after the $n$th one \cite{Feller}, 
and is well investigated. 
Taking $\tilde\psi(u) \simeq 1-Au^\alpha$ in Laplace domain and using the convolution theorem results 
in $\tilde{p}_n(u) = Au^{\alpha-1}\exp\left[n \ln (1-Au^\alpha) \right] $. Laplace inversion yields
\begin{equation}
p_n(t) = \frac{1}{\alpha}\frac{t}{A^{1/\alpha} n^{1+1\alpha}} l_{\alpha,1}\left[\frac{t}{A^{1/\alpha}n^{1/\alpha}} \right] . \label{ml-dist}
\end{equation}
$l_{\alpha,1}(t)$ denotes the one-sided L\'evy density whose Laplace transform is given by $\exp[-u^\alpha]$ \cite{Feller}. 
Hence, with $\langle n_t \rangle \sim t^\alpha(A\Gamma(1+\alpha)) $ the ensemble average of Eq. (\ref{tamsd-pr}) becomes
\begin{equation}
\langle \overline{\delta^2} \rangle = v_0^2 \Delta^2 - \frac{2v_0^2}{3A\Gamma(1+\alpha)}t^{\alpha-1}\Delta^3 \label{EAtamsd2}
\end{equation}
which clearly differs from Eq. (\ref{EAtamsd}). Eq. (\ref{tamsd-pr}) and hence (\ref{EAtamsd2})
describes a special case of the
sojourn time distribution Eqs. (\ref{psioft}), (\ref{psiofu}) fulfilling the relations
$t^{1-\alpha}\gg \Delta/(A\Gamma(1+\alpha))$ and $\Delta\leq 1$ so that the first ballistic term in 
Eq. (\ref{EAtamsd}) remains larger than second term. In contrast, Eq. (\ref{EAtamsd}) requires
$\Delta\gg 1$.
From Eq. (\ref{tamsd-pr}) we find that the quantity
$(v_0^2\Delta^2-\overline{\delta^2})$ and the amount of switching events within $t$ are proportional,
\begin{eqnarray}
(v_0^2\Delta^2-\overline{\delta^2}) 
&=&
\frac{2}{3}\frac{ \Delta^3}{t} n_t
\end{eqnarray}
Therefore, in terms of a new variable 
\begin{equation}
\xi = \frac{v_0^2 \Delta^2 - \overline{\delta^2} }{v_0^2 \Delta^2 - \left\langle \overline{\delta^2}\right\rangle } 
= \frac{n_t}{\langle n_t \rangle} \label{xi}
\end{equation}
and  using Eq. (\ref{ml-dist}) the rescaled distribution of the TAMSD becomes
\begin{eqnarray}
p(\xi) &=& \frac{\Gamma^{1/\alpha}(1+\alpha)}{\alpha \xi^{1+1/\alpha}}
l_{\alpha,1}\left[\frac{\Gamma^{1/\alpha}(1+\alpha)}{\xi^{1/\alpha}} \right] \, .\label{ML-th}
\end{eqnarray}
This PDF is the density of the Mittag-Leffler distribution, a distribution already encountered in the context of
TAMSD fluctuations in the subdiffusive continuous time random walk \cite{HeBur08}, \cite{Aki11}.
Fig. \ref{ML-figa} shows the PDF (\ref{ML-th}) and the respective results for simulations of the 
L\'evy-Walk for two different values of $\alpha$, but for large $\Delta$.
It is interesting to note that $\langle v_0^2\Delta^2-\bar{\delta^2}\rangle$
differs for small and large $\Delta$ regimes, 
however the distribution of the rescaled variable (\ref{xi}) does not depend on $\Delta$. 

\begin{figure}[h]\centering{
\includegraphics[width=.5\textwidth]{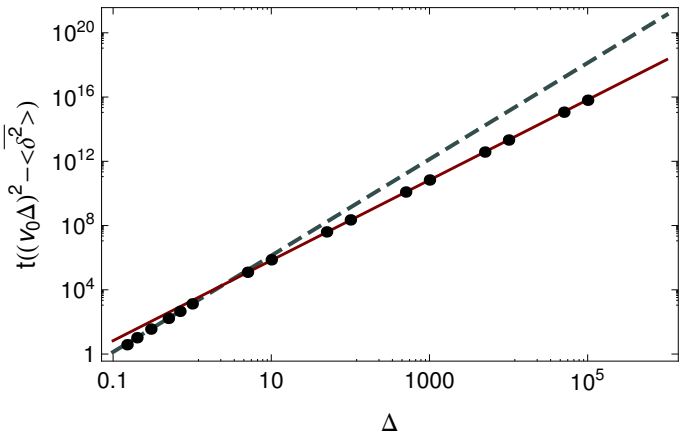}}
\caption{\label{ML-cross} $t (v_0^2\Delta^2-\left\langle\bar{\delta^2}\right\rangle)$ versus $\Delta$, $\alpha=0.5$.
The small--$\Delta$ region is sensitive to the shape of $\psi(\tau)$. 
Lines indicate theory Eq. (\ref{EAtamsd2}) (dashed)
matching numerical data (dots) at small $\Delta$, and Eq. (\ref{EAtamsd}) (solid) for large $\Delta$. 
Note that the crossover takes place 
in the region of the small-time cutoff of the sojourn time PDF Eq.(\ref{psioft}), $\Delta_{cr}\approx 1$.
Sample size $10^4$, $v_0=1$, $t=10^7$.}
\end{figure}

\paragraph{Crossover to the large $\Delta$ regime:}
The ensemble average of the fluctuations of $ (v_0^2\Delta^2-\bar{\delta^2})$ is given by 
Eq. (\ref{EAtamsd}) for large $1\ll\Delta\ll t$.
In the present case where $\psi(\tau)=0$ at short times $\tau<1$, the behavior of the 
ensemble averaged TAMSD at $\Delta\ll 1$ is given by Eq. (\ref{EAtamsd2}).
This behavior at small $\Delta$ constitutes the lower bound for more general $\psi(\tau)$ with 
arbitrary shape at small $\tau$. 
However, the fluctuations of the time averages Eq. (\ref{ML-th})
are governed only by the tail of the persistence PDF $\psi(\tau)$ and are therefore the same for small and large $\Delta$. 
Hence we can write 
\begin{eqnarray}
&& v_0^2\Delta^2 -  \bar{\delta^2}  = \chi^2 n_t, \nonumber\\
&& \chi^2 = \left\lbrace
\begin{array}{c  l}
\frac{2v_0^2 \Delta^3}{3t}  & \hspace{.3cm} \Delta \ll 1 \\
\frac{2v_0^2 \sin\pi\alpha \Delta^{3-\alpha}}{\pi\alpha(6-11\alpha +6\alpha^2- \alpha^3) A\Gamma(1+\alpha) t}  & \hspace{.3cm} \Delta \gg 1
\end{array} 
\right. .\label{ChiSq}
\end{eqnarray}
Here $\chi^2$ gives the deterministic part that governs the ensemble mean of the shifted TAMSD, while the full
fluctuations enter via $n_t$, compare Eqs. (\ref{xi}), (\ref{ML-th}).
In Fig. \ref{ML-cross} we plot $t(v_0^2\Delta^2 - \left\langle \bar{\delta^2} \right\rangle)$ versus $\Delta$.
Simulational results match the theoretical short time as well as long time behaviors, 
Eqs. (\ref{EAtamsd2}) and (\ref{EAtamsd}), respectively.
The crossover takes place in the region of the cutoff of the sojourn time PDF 
$\psi(\tau)$ at small times, i.e.  at $\Delta_{cr}\approx 1$.


\begin{figure}[h]\centering{
\includegraphics[width=.45\textwidth]{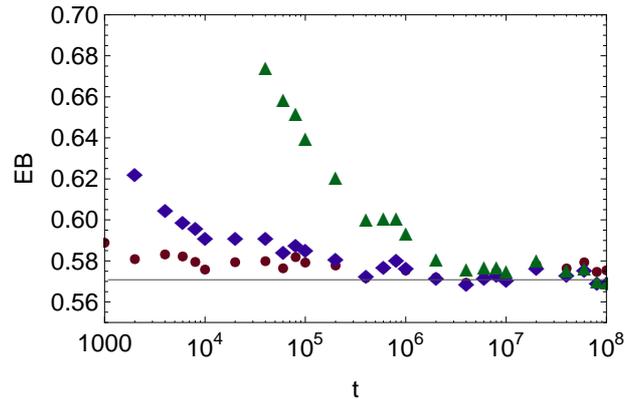}}
\caption{\label{EB-b}  EB-parameter of $\xi$ for $\alpha=0.5$. Red circles indicate $\Delta=0.5$, blue diamonds $\Delta=100$, 
green triangles $\Delta=5000$. The grey solid line indicates the theoretical value $\mbox{EB}=0.571$, Eq. (\ref{EBxi}). 
Sample size $5000$.}
\end{figure}

Finally we demonstrate numerically that the above distributions are 
indeed the limiting distributions at large times. 
For this purpose, we calculate the ergodicity breaking (EB) parameter \cite{HeBur08} for the 
shifted TAMSD $\xi$
\begin{eqnarray}
\mbox{EB} &=& \lim_{t\to\infty} 
\frac{\left\langle \xi^2\right\rangle - \left\langle \xi\right\rangle^2}
{\left\langle \xi\right\rangle^2} 
= \frac{2\Gamma^2(1+\alpha)}{\Gamma(1+2\alpha)}-1 ,\label{EBxi}
\end{eqnarray}
where we used Eq. (\ref{ML-th}).
Numerics for $\alpha=0.5$ show that the EB-parameter for $\xi$ tends indeed to the predicted finite value 
$\mbox{EB}=0.571$ (Fig. \ref{EB-b}), i.e. the variable $\xi$ remains distributed 
according to Eq.(\ref{ML-th}).

Note that, however, the $\mathrm{EB}$-parameter for the original (not shifted) 
TAMSD $\overline{\delta^2}$ slowly tends to zero for nonzero $\Delta$ as $t^{-2(1-\alpha)}$.
Hence, for the ballistic L\'evy walk, non-ergodicity in the sense of the distribution of time averages
does not find its expression in the (decaying) fluctuations of the TAMSDs $\overline{\delta^2}$ themselves,
but in the (persisting) fluctuations of the shifted and rescaled variable $\xi$.

\section{L\'evy flight TAMSD}

Before we turn to the behavior of the TAMSD of the L\'evy walk in the enhanced diffusion regime, 
let us illustrate the situation in the related L\'evy flight.
We present a rather illustrative than rigorous argument, to avoid complicated math. 
A more general and rigorous treatment can be found in Ref. \cite{BurWer10}.
The L\'evy flight is a random walk process where at each renewal the displacement $x$ of the walker 
is drawn according to a jump PDF $\lambda(x)$,
but in contrast to the normal random walk the jump PDF lacks the second moment.
With a unit time span passing between consecutive renewal events, the
number of jumps acts as the (discrete) time variable $t$.
Hence consider the coordinate of a L\'evy flight after $t$ steps as a sum of independent identically distributed (i.i.d.)
random variables or displacements $x_i$
\begin{eqnarray}
X_t &=& \sum_{i=1}^t x_i\, .  \nonumber
\end{eqnarray}
Let the $\left\lbrace x_i \right\rbrace $ be distributed according to a two-sided symmetric 
distribution $\lambda(x)=\lambda(-x)$ falling off as a power law for large $|x|$. 
\begin{equation}
\lambda(x) \propto \frac{A_0}{2} |x|^{-1-\alpha}\;,\hspace{1cm} 1<\alpha<2, \label{jumplpdf}
\end{equation}
In particular, for our simulations we used Eq. (\ref{jumplpdf}) with $A_0 = \alpha$ for $|x|\geq 1$, 
and $\lambda(x)=0$ for $|x|<1$.
Then, the distribution of the sum $X_t$ will yield a two-sided L\'evy law for large $t$,
$L_{\alpha,0}( X_t/(t A_0 \Gamma(1-\alpha)/\alpha)^{1/\alpha})$, according to the generalized 
central limit theorem \cite{Bou90,Feller}. 
The function $L_{0,\alpha}(x)$ is defined as the inverse Fourier
transform of 
\begin{eqnarray}
\hat{L}_{\alpha,0}(k) &=& \exp \left[-|k|^{\alpha}\right]
\end{eqnarray}
with $k$ being the Fourier variable \cite{Feller}.
This Green function of the L\'evy flight has a similar behavior as the central part of the L\'evy walk 
Green function when $1<\alpha<2$, as will become obvious in the next section.

\begin{figure}[h]
\centering{
{\includegraphics[width=.45\linewidth]{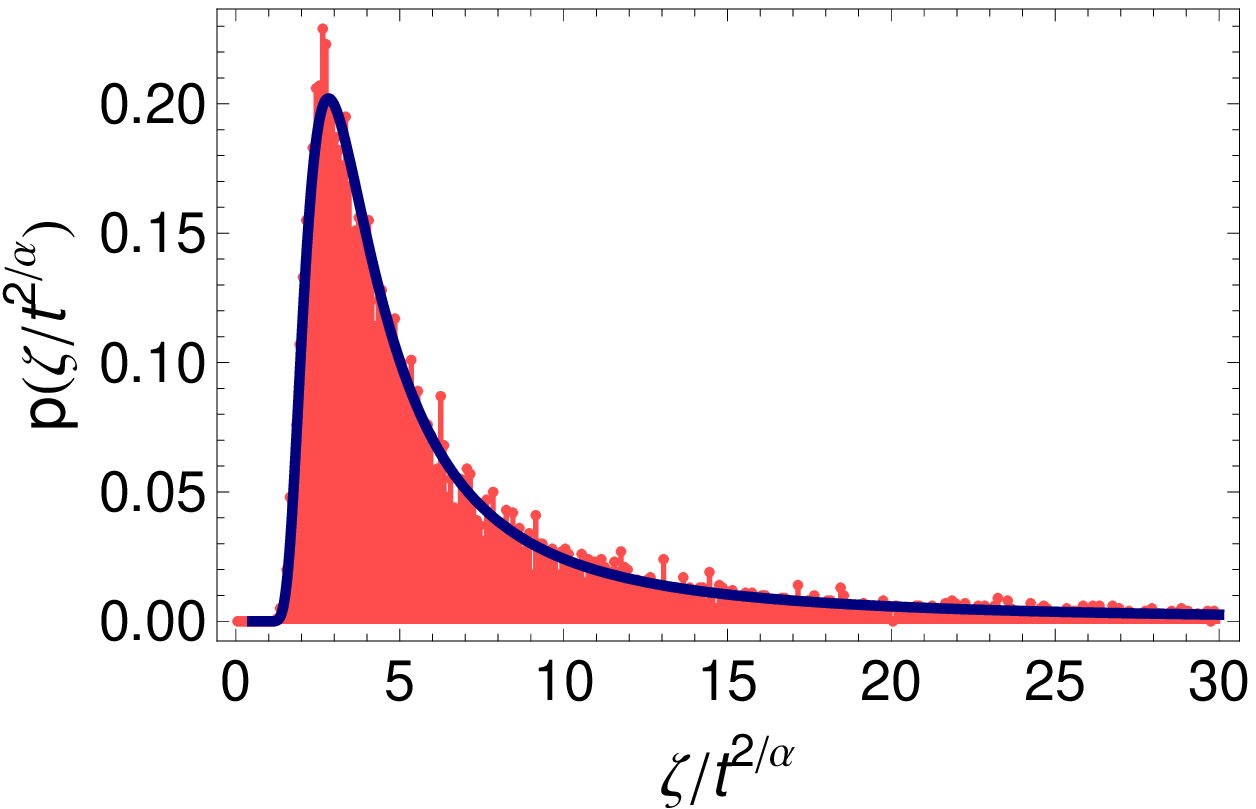}}
\hspace{.1cm}
{\includegraphics[width=.45\linewidth]{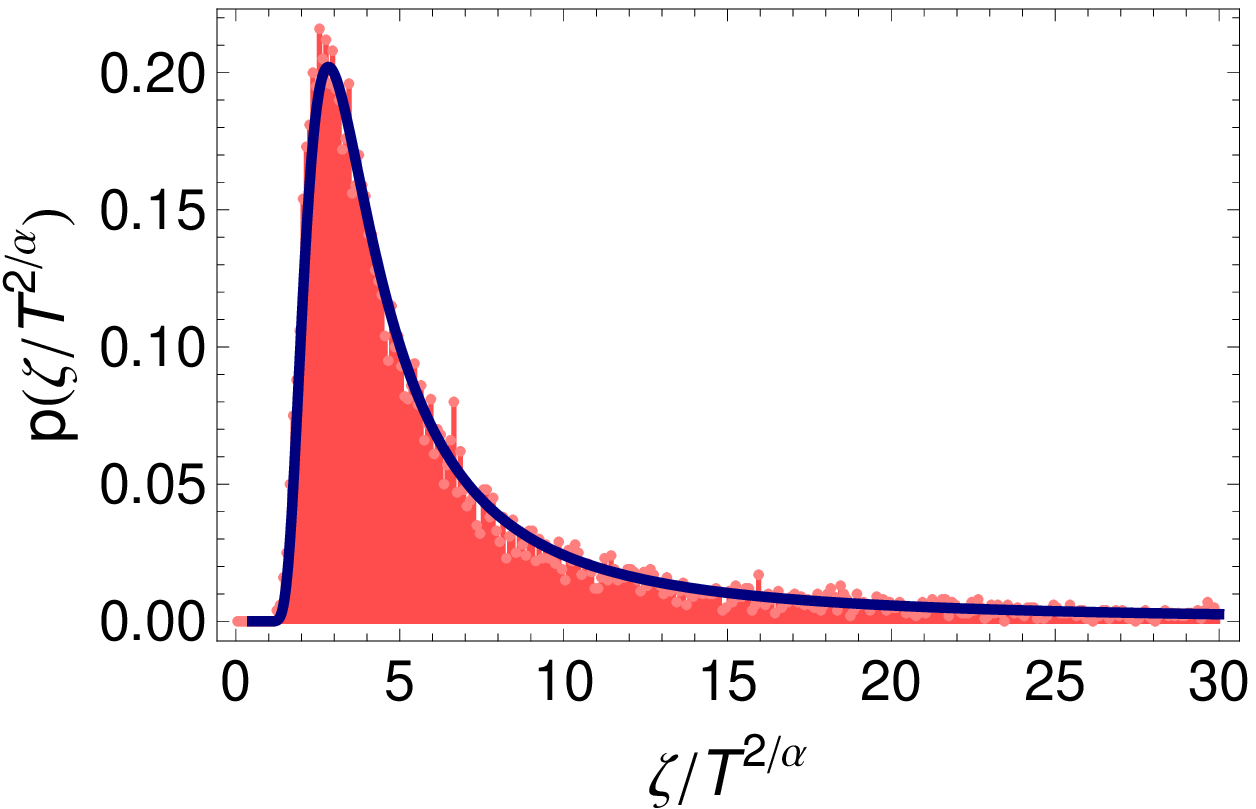}}
}
\caption{\label{ta} L\'evy flight simulation results for the distribution of 
$\frac{\zeta}{t^{\frac{2}{\alpha}}}$ for $\alpha = 1.5$ and $K_\alpha^{LF}=\Gamma(1-\frac{\alpha}{2})$, 
$\Delta =1$, $t=10^5$ (left)
and $\Delta = 100$, $t=10^6$ (right),
and theoretical predictions according to Eq. (\ref{LFtamsd}) 
(solid curves).}
\end{figure}

The time-averaged mean squared displacement (TAMSD) is defined as \cite{BurWer10}
\begin{eqnarray}
\overline{\delta^2} &=& \frac{1}{t-\Delta} \sum_{i=1}^{t-\Delta} \left[ X_{i+\Delta} - X_i\right]^2 \nonumber \\
&=& \frac{1}{t-\Delta} \sum_{i=1}^{t-\Delta} \left(\sum_{k=i}^{i+\Delta} x_k\right)^2
\end{eqnarray}
where the integer $\Delta$ is the lag time. We have
\begin{eqnarray}
\overline{\delta^2} &=&
\frac{1}{t-\Delta} \sum_{i=1}^{t-\Delta}\left(
\sum_{k=i}^{i+\Delta} x_k^2 + \sum_{k=i}^{i+\Delta}\sum_{j\neq k}^{i+\Delta} x_k x_j
\right) 
\end{eqnarray}
The mixed terms $\sum_{k=i}^{i+\Delta}\sum_{j\neq k}^{i+\Delta} x_k x_j$ on average cancel out
for large enough $\Delta$, hence we omit them. Moreover we assume $1\ll\Delta\ll t$ so that
\begin{eqnarray}
\bar\delta^2 &\simeq&
\frac{1}{t-\Delta}
\sum_{i=1}^{t-\Delta}\sum_{k=i}^{i+\Delta} x_k^2 \nonumber \\
&\simeq&\frac{1}{t} 
\sum_{i=1}^{t}\sum_{k=i}^{i+\Delta} x_k^2 \nonumber\\
&{\buildrel d\over\approx}& \frac{\Delta}{t} \sum_{k=1}^{t} x_k^2
\label{tamsd}
\end{eqnarray}
We find for the distribution of the $y=x^2$
\begin{eqnarray}
p(x^2) &=& p(x)\left|\frac{dx}{dy}\right|
\propto \frac{A_0}{2} y^{-1-\frac \alpha 2}\, .
\end{eqnarray}
Note that the transition from $x$ to the positive valued $y$ results in a factor $2$ in the normalization.
The large $(y = x^2)$ asymptotics can be obtained in Laplace domain, using the Tauberian theorem:
\begin{equation}
\tilde p(u_y) \simeq 1 - \frac{A_0}{\alpha} \Gamma(1-\frac{\alpha}{2}) u_y^{\frac{\alpha}{2}}\, .
\end{equation} 
Hence, the sum over these $x_k^2 = y_k$ in Eq. (\ref{tamsd}) is a sum over positive i.i.d. random variables 
distributed according to a PDF with a power law tail of exponent $-1-\alpha/2$.
Hence, due to the generalized central limit theorem we find in the large $t$ limit \cite{Bou90}, \cite{Feller}the PDF of $\zeta=t\bar{\delta^2}/\Delta$:
\begin{equation}
p\left(\zeta \right) = \frac{1}{\left(K_\alpha^{LF}t\right)^{\frac{2}{\alpha}} }
L_{\frac{\alpha}{2},1} 
\left\lbrace \frac{\zeta}{ \left(K_\alpha^{LF}t\right)^{\frac{2}{\alpha}} } \right\rbrace \, . \label{LFtamsd}
\end{equation}
where $K_\alpha^{LF}=\frac{A_0}{\alpha} \Gamma(1-\frac{\alpha}{2})$ and $L_{\alpha/2,1}(\cdot)$ 
is the one-sided L\'evy PDF given by $\exp\left[ -u^{\alpha/2}\right] $ in Laplace domain \cite{Feller}.
A similar result was obtained in \cite{BurWer10} though a slightly different scaling was reported.
Fig. \ref{ta} shows the PDF Eq. (\ref{LFtamsd}) obtained with {\scshape Mathematica}, and the corresponding simulational results 
which perfectly match the theory.
 
\begin{figure}[h]
\centering{
{\includegraphics[width=.45\textwidth]{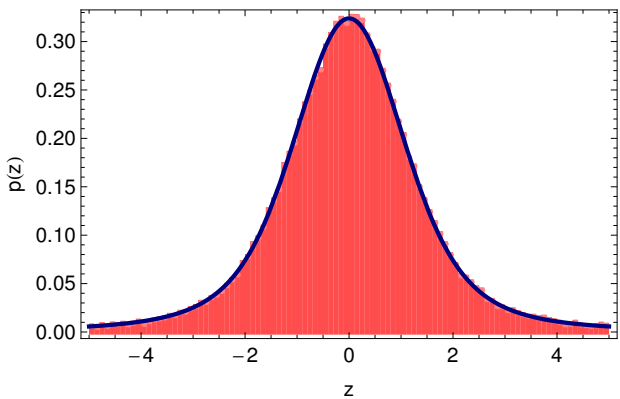}}
\caption{\label{Lv-SB} For the L\'evy walk with $1<\alpha<2$ the central part 
of the particle distribution (vs. the scaling variable $z=x/t^{1/\alpha}$) is a 
symmetric L\'evy distribution Eq. (\ref{scPos}) (solid line). 
Simulation (histogram) for $\alpha=3/2$, $t=10^6$, $v_0=1$, 
$K_\alpha$ see Eq. (\ref{Kalph}), sample size $10^5$.}}
\end{figure}

Hence, this example illustrates that the TAMSD of L\'evy Flights with jump length distributions without 
second moment is a one-sided L\'evy density lacking the first moment. As will be shown later,
the TAMSD distribution for the corresponding L\'evy walk is competely different despite the similarity 
in the central part of the propagator, see Fig. \ref{Lv-SB}.

\section{Enhanced diffusion regime}

In this section we consider the regime of sojourn times distributed according to a PDF with existing mean $\langle \tau \rangle$, 
but diverging second moment, i.e. Eq. (\ref{psioft})
with $1<\alpha< 2$. The expansion in Laplace domain is hence
\begin{equation}
\tilde{\psi}(u) \simeq 1 - \left\langle \tau \right\rangle u + A u^\alpha .\label{SBLapPsi}
\end{equation}
In our case Eq. (\ref{psioft}) the average sojourn time is 
$\left\langle \tau \right\rangle = \alpha/(\alpha-1)$ and $A = |\Gamma(1-\alpha)|$.
\subsection{Particle position distribution}
Again, the particle position is given by the integral over the velocities $v_0(t_+ - t_-)$. 
The MSD is well known \cite{GodrLuck01}:
\begin{eqnarray}
\left\langle x^2 (t)\right\rangle &\simeq& 
\frac{2 v_0^2 A}{\left\langle \tau \right\rangle} 
\frac{(\alpha-1)}
{\Gamma(4-\alpha)} 
t^{3-\alpha}, \label{SBmsd}
\end{eqnarray}
as well as the propagator which is given by a  two-sided L\'evy-distribution for the central part where $|x|\ll v_0 t$,
\begin{eqnarray}
p(x,t) \simeq \frac{1}{\left( K_\alpha t\right) ^{1/\alpha}} L_{\alpha,0} 
\left(\frac{ x}{\left( K_\alpha t\right) ^{1/\alpha}} \right) .
\label{scPos}
\end{eqnarray}
where $\hat {L}_{\alpha,0}(k) = \exp\left[ -k^{\alpha}\right]$ and 
\begin{equation}
K_{\alpha}=\frac{v_0^2 A (\alpha-1)}{\langle\tau\rangle\Gamma(4-\alpha)}. \label{Kalph}
\end{equation}

Due to the finite velocity the L\'evy walk propagator exhibits a cutoff so that $p(x,t) = 0$ at $|x|>v_0t$,
similarly to other L\'evy walk models \cite{ZumKlaf90}, \cite{ShleKlaf89}. 
Moreover, we expect those particles that have never changed their direction of motion to form a delta-peak 
at the edge of the cutoff \cite{BarFleuKlaf00}.
In the following we will calculate the time averaged mean squared displacement (TAMSD) of the L\'evy Walk 
described above. 

\subsection{Ensemble average of $\overline{\delta^2}$}

It is important to note that also in the subballistic regime $1<\alpha<2$, the velocity correlation 
is governed by the persistence probability
$p_0(t)$ to stay in one state for a time $t$:\newline
For sojourn time PDFs $\psi(\tau)$ with existing mean 
the corresponding forward recurrence time PDF $\psi_{f,t_1}(\tau_f)$ 
reaches stationarity at large $t_1$ and obeys the limiting distribution \cite{GodrLuck01}
\begin{eqnarray}
\lim_{t_1\to\infty}\psi_{f, t_1}(\tau_f) &=& \frac{1}{\left\langle \tau \right\rangle } \int_{\tau_f}^\infty \psi(\tau)d\tau \simeq 
\frac{A}{\left\langle \tau \right\rangle \left|\Gamma(1-\alpha)\right|} \tau_f^{-\alpha}. \nonumber
\end{eqnarray}
This first waiting time in turn has no mean and therefore $p_0$ dominates $\langle v(t_1)v(t_1+\Delta)\rangle$.
Again, with Eqs. (\ref{genCvv}), (\ref{genCvva}), $\langle v(t_1)v(t_1+\Delta)\rangle = v_0^2 p_0(t_1,t_1+\Delta)$ 
holds for the velocity correlation function for $t_1$ and $\Delta$ both large.
The $p_0$ for the present process is well known, so that following e.g. the procedure presented in \cite{GodrLuck01} 
we have
\begin{eqnarray}
&& \langle v(t_1)v(t_1+\Delta)\rangle 
= \frac{v_0^2A}{(\alpha-1)\left\langle \tau\right\rangle |\Gamma(1-\alpha)|}
t_1^{1-\alpha} \times \nonumber \\
&& \left( \frac{t_1}{t_1+\Delta}\right)^{\alpha-1}
\left( \left(1-\frac{t_1}{t_1+\Delta} \right)^{1-\alpha}  - 1 \right).  \label{CvvGL}
\end{eqnarray}
In the equilibrated regime (or stationary state) $t_1 \gg \Delta$, $\langle v(t_1)v(t_1+\Delta)\rangle$ 
becomes independent of time $t_1$ so that
\begin{eqnarray}
\langle v(t_1)v(t_1+\Delta)\rangle_{\mathrm eq} &\simeq& \frac{v_0^2 A}{\left\langle \tau\right\rangle (\alpha-1) |\Gamma(1-\alpha)|} \Delta^{1-\alpha}. \label{statCvv}
\end{eqnarray}
By integrating Eq. (\ref{CvvGL}) as in Eq. (\ref{Cxx0}) we obtain the position autocorrelation
\begin{eqnarray}
&&\left\langle x(t_1) x(t_1+\Delta)\right\rangle =
 \frac{A v_0^2}{\left\langle \tau \right\rangle \Gamma(4-\alpha)} t_1^{3-\alpha} \times \nonumber \\
&&\left[ -\left( y\right) ^{3-\alpha} + \left(1+y\right)^{3-\alpha}  
+ (\alpha-3)\left(1+y\right)^{2-\alpha} + \alpha  \right] 
\label{CxxGL}
\end{eqnarray}
where $y=\Delta/t_1$. 
Our simulations have shown that this estimation reproduces the large $t_1$ behavior of the position correlation
$\langle x(t_1)x(t_2)\rangle$ quite well.
For $\Delta\to 0$ the behavior of the MSD Eq. (\ref{SBmsd}) is reproduced.

\begin{figure}[h]
\centering{
{\includegraphics[width=.8\linewidth]{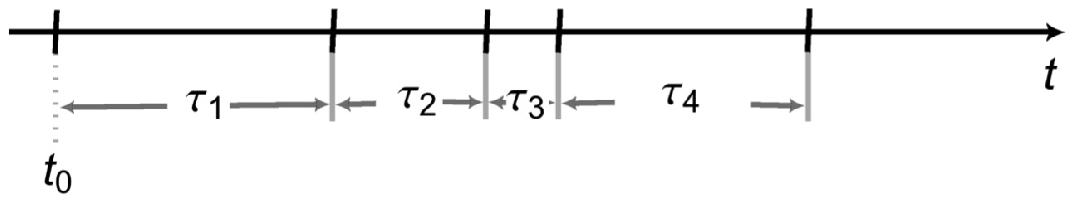}}\\
\vspace{.1cm}
{\includegraphics[width=.8\linewidth]{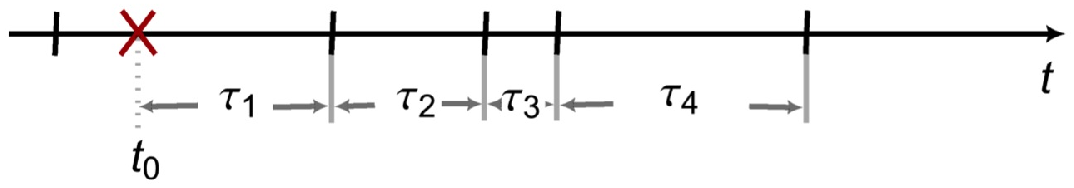}}
}
\caption{\label{ini} Sketch of the process starting at the beginning of the measurement $t_0$ 
(upper panel)
and  the equilibrium process starting at $t=-t_{ini}$ with $t_{ini}\ll \langle \tau\rangle$. 
The measurement begins at a 
time $t_0$ between two renewal events (lower panel). 
Renewal events are indicated by black ticks, the start of the measurement $t_0$ is marked by a cross.}
\end{figure}

Note that in the subballistic case the MSD for a process starting with the beginning of the measurement 
differs from the MSD for a process that started a long time before the beginning of the measurement time $t_0$, 
as was found earlier in the context of a stochastic collision model \cite{BarFleu97}. This behavior is
due to the predominant role of the persistence probability $p_0$ in the correlation functions discussed above.
In what follows the MSD for the process that started long before $t_0$ will be called 
the equilibrium MSD $\left\langle x^2 \right\rangle_{\mathrm eq}$, alluding to the fact that this process has no memory 
of its starting time. The situation is sketched in Fig. \ref{ini}.
It is important to introduce the equilibrium MSD at this point for the following reason: 
Since the time averaging procedure comprises averaging over all continuously shifted time lags, 
and not only over those starting at a switching event, there is an inherent averaging over disorder.
The equilibrium MSD accounts for this averaging over disorder and is therefore the natural 
ensemble averaged quantity to later compare the time averaged MSD to. Note also that such a definition of
an equilibrium MSD is not possible in the ballistic case since there a stationary state 
does not exist -- the respective MSD would never become independent of the time difference 
between start of the process in the past and the actual start of the measurement.
 
\begin{figure}[h]
\centering{
{\includegraphics[width=.45\textwidth]{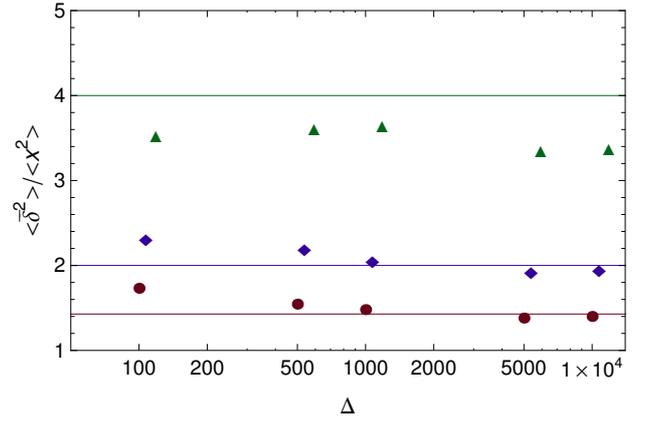}}
\caption{\label{factor} Ratio $\left\langle \overline{\delta^2}\right\rangle/\left\langle x^2 \right\rangle$ for
$\alpha = 1.25$ (triangles), $\alpha=1.5$ (diamonds) and $\alpha=1.7$ (circles); $t=10^7$. 
The theory predicts $4$, $2$ and $1.43$ for this ratio (solid lines).
}}
\end{figure}

Using Eq. (\ref{statCvv}) and
$\left\langle x^2 \right\rangle_{\mathrm eq} = 2\int_0^t dt_1 \int_0^t dt_2 \langle v(t_1)v(t_2)\rangle_{\mathrm eq} $
we obtain the known equilibrium MSD \cite{BarFleu97}
\begin{equation}
\langle x^2 \rangle_{\mathrm eq} = 
v_0^2 \frac{A}{\langle \tau\rangle }
\frac{2}{\Gamma(4-\alpha)} t^{3-\alpha} \label{statmsd}
\end{equation}
For the TAMSD we use again the definition Eq. (\ref{tamsd0})
and write Eq. (\ref{TAMSD2}) for the respective ensemble average.
To obtain a description of the ensemble averaged TAMSD, we insert the 
autocorrelation function Eq. (\ref{CxxGL}) and the MSD Eq. (\ref{SBmsd}) into Eq. (\ref{TAMSD2}). 
We thus have
\begin{eqnarray}
&&\langle \overline{\delta^2}\rangle  \simeq \frac{v_0^2}{(t-\Delta)} \frac{2A}{\langle \tau\rangle \Gamma(5-\alpha)}
\times \nonumber\\
&& \left[ 
-\left( t-\Delta\right)^{4-\alpha} + t^{4-\alpha} -\Delta^{4-\alpha} \right.\nonumber \\
&&\left. + (4-\alpha)t\Delta^{3-\alpha} - (4-\alpha)t^{3-\alpha}\Delta
\right] \label{taDel0}
\end{eqnarray}
which becomes in leading orders by expansion for $\Delta/t$ small:
\begin{eqnarray}
\langle \overline{\delta^2}\rangle  &\simeq& 
\frac{2Av_0^2}{\langle \tau\rangle\Gamma(4-\alpha)} \Delta^{3-\alpha}
- \frac{Av_0^2}{\langle \tau\rangle\Gamma(3-\alpha)} \Delta^2 t^{1-\alpha}   \label{taDel}
\end{eqnarray}
Note that this result for $\langle \bar \delta^2\rangle$ complies with the time dependence of
the equilibrium ensemble average $\left\langle x^2 \right\rangle_{\mathrm eq}$, Eq. (\ref{statmsd}), and hence 
differs from the MSD (\ref{SBmsd}) by lacking a factor:
\begin{equation}
\lim_{t\to\infty} \frac{\overline{\delta^2}}{\langle x^2\rangle} = \frac{1}{\alpha-1}.
\end{equation}
Numerical simulations also indicate that these two averages appear to differ by a factor (Fig. \ref{factor}).
Further numerical evidence for this behavior was found in \cite{Akimoto12}, \cite{Ralf13}.
Especially for small $\alpha$ the convergence of $\langle \bar \delta^2\rangle$ is extremely slow.
Moreover, in our simulations the large time behavior of the $\left\langle x^2 \right\rangle $ is not represented very accurately at the corresponding relatively small times (up to $10^4$).

\subsection{Fluctuations of the TAMSD}

\begin{figure}[h]
\centering{
{\includegraphics[width=.45\textwidth]{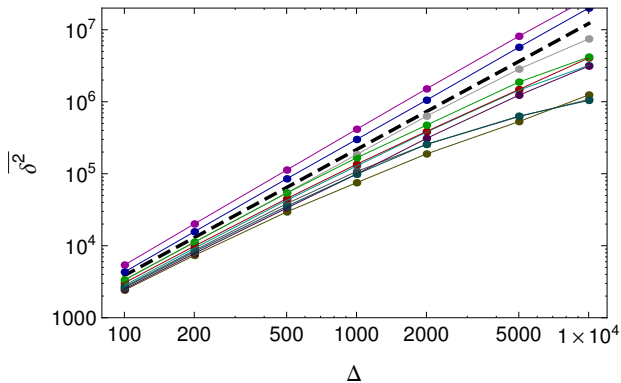}}
\caption{\label{traj125} TAMSDs of particle trajectories in dependence on the lag time $\Delta$, 
$t=10^{5}$;
$\alpha = 5/4$. For finite measurement times $t$ fluctuations in $\overline{\delta^2}$ are observed. 
Larger times $t$ and smaller lag times $\Delta$ result in smaller fluctuations.}}
\end{figure}

The TAMSDs of trajectories measured up to a certain observation time appear to be distributed.
Fig. \ref{traj125} shows the TAMSD evolution of some sample trajectories.
The fluctuations of the TAMSDs decrease with increasing total measurement time. However, 
since in experiments the observation time is always finite, these fluctuations may play a role in practice.
In our simulations for example we find large fluctuations among the 
$\overline{\delta^2}$ for $\alpha=1.25$ and $\Delta=100$, $t=10^5$ (see Fig. \ref{traj125}).

\begin{figure}[h]
\centering{
{\includegraphics[width=.45\linewidth]{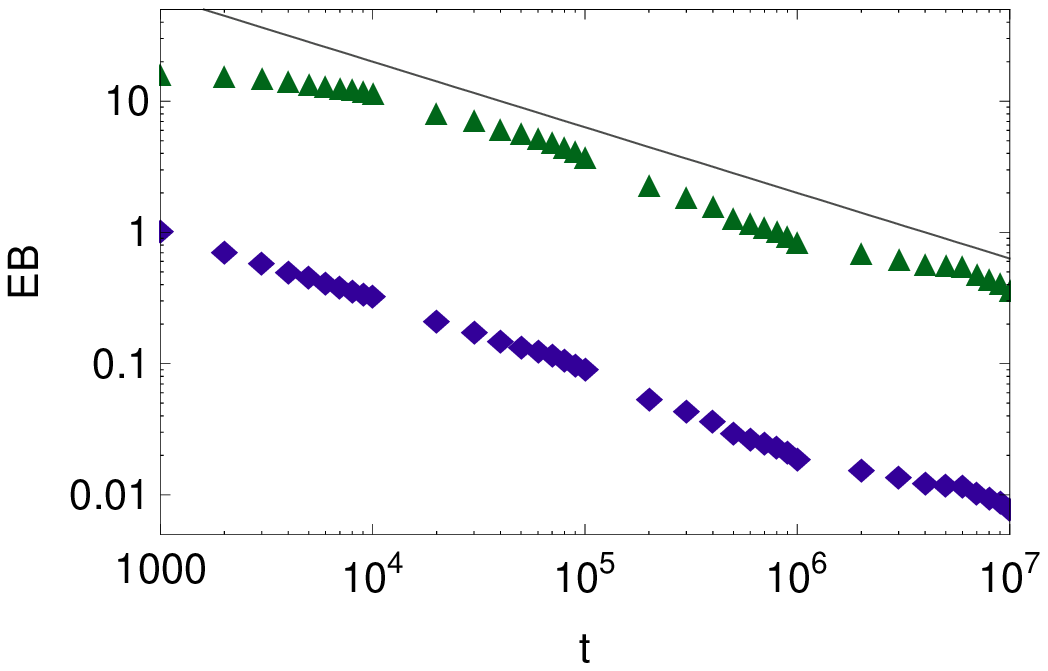}}
\hspace{.05cm}
{\includegraphics[width=.45\linewidth]{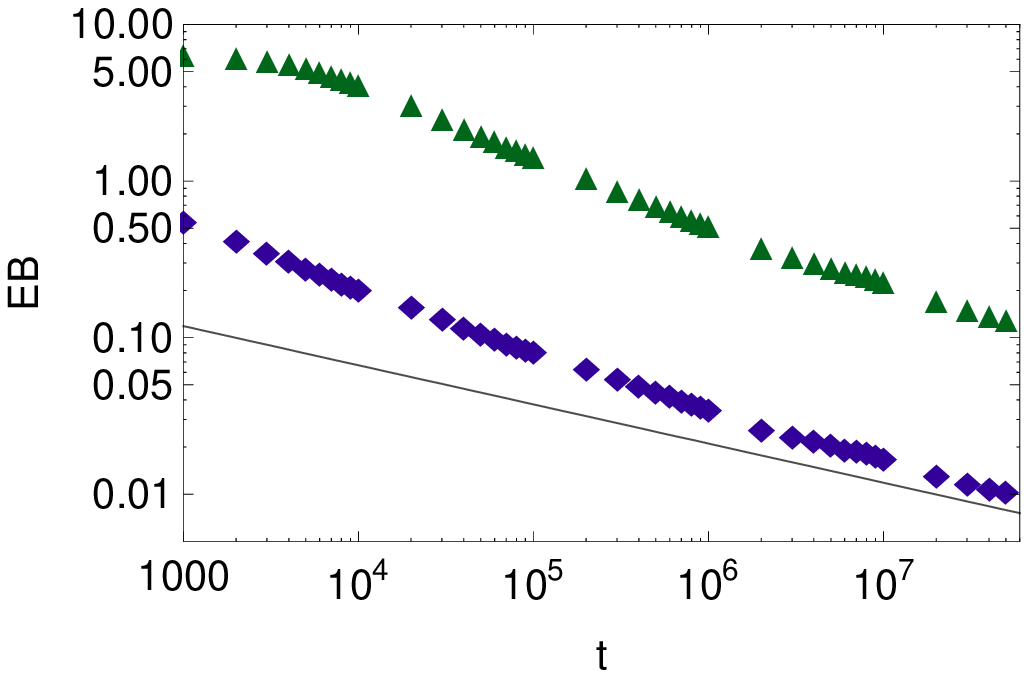}}}
\caption{\label{EB-SB} EB-parameter Eq. (\ref{EBSB}), for $\alpha=3/2$ (left) and $5/4$ (right). Blue diamonds indicate $\Delta=100$, 
green triangles $\Delta=5000$, the grey solid line indicates the slope $(1-\alpha)$.}
\end{figure}

Although the propagators of the L\'evy walk and flight look very similar in the central part $|x|<v_0 t$,
unlike the flight case simulations suggest that the TAMSD distribution for the L\'evy walk cannot be expressed 
by a L\'evy distribution of stability index $\alpha/2$.
For the Ergodicity Breaking (EB) parameter of the subballistic L\'evy Walk regime,
\begin{eqnarray}
EB &=& \lim_{t\to\infty} 
\frac{\left\langle \left( \overline{\delta^2}\right)^2\right\rangle - \left\langle \overline{\delta^2}\right\rangle^2}
{\left\langle \overline{\delta^2}\right\rangle^2}, \label{EBSB}
\end{eqnarray}
we find a steady decay with $t$, which is yet very slow with an exponent of roughly $(1-\alpha)$, 
to the value zero indicating that the width of the distribution tends to zero (Figs. \ref{EB-SB}, for $\alpha=1.5\; , 1.25$).
Hence, the TAMSDs do not remain distributed in the limit of very long times in the subballistic regime. 
Such a very slow decay to ergodic behavior is not a distinct feature of the enhanced phase of the 
L\'evy walk model but can also be found for completely different ergodic systems such as relaxation of 
confined fractional Brownian motion \cite{Jeon12}.

In contrast to the flight case, the width of the $\overline{\delta^2}$--distribution for the L\'evy walk at finite times 
always exists due to finite velocity.
Simulations suggest that it increases with the lag time $\Delta$ as $\Delta^4$.
The width of the TAMSD distribution hence appears to have the same $\Delta$-dependence as a ballistic motion.
However, it is not quite clear whether this dependence represents the large time behavior 
of the distribution of the TAMSDs,
or whether it is an artefact of the ballistic tails of the propagator due to the extremely long transients.

\section{Summary}

In this article we have shown that the shifted time averaged MSD of the ballistic L'evy walk is described
by the Mittag-Leffler distribution, similar to the distribution of the TAMSD in the sub-diffusive 
continuous time random walk (CTRW). This distribution
describes the fluctuations of the time averages and is universal.  The TAMSD averaged
over an ensemble of trajectories $\langle \overline{\delta^2} \rangle$ is not equal to the ensemble
average $\langle x^2 \rangle$ as already pointed out by Akimoto \cite{Akimoto12}. 
Interestingly $\langle \overline{\delta^2} \rangle- (v_0 \delta^2)$
exhibits two behaviors valid for $\Delta<1$ and $\Delta>1$, and it would be interesting to see 
if a similar cross-over takes place in other models such as the sub-diffusive CTRW. 
For L\'evy flights the TAMSDs are random with a PDF given by the one sided L\'evy PDF, 
which is in agreement with rigorous results (though note that our coefficients are different than those reported
in \cite{BurWer10}, possibly due to a typo). 
In the enhanced diffusion regime, the PDF of the particle position of the L\'evy walk is similar
to the L\'evy flight case, at least in its center. However, the TAMSD of the two models is vastly
different, and for L\'evy walks no fluctuations are found for $\overline{\delta^2}$. 
This indicates that the TAMSD is controlled by rare events, 
since the tails of the mentioned distributions are where one finds differences between the models. 
Thus taking into consideration finite velocity (like in the L\'evy walk model) is crucial for 
our understanding of the ergodic properties of these processes. 
In the future it might be worth while checking the time averages
of lower order moments, since they might exhibit behavior very different the second moment
considered here. We note that for finite times the fluctuations of TAMSDs are large.
Consequently, in the laboratory where experiments are made
for finite time the process may seem non ergodic, but this is only a finite time effect. Moreover,
in this sub-ballistic case $\langle \overline{\delta^2} \rangle$ is equal to
the equilibrium MSD $\langle x^2 \rangle_{\mathrm eq}$, but not to $\langle x^2 \rangle$. 
If one wishes to compare time and ensemble averages, the conclusion on equality of these two averages
will depend on how the ensemble is prepared. To attain ergodicity we need to start the ensemble in a stationary
state, which is not so surprising. The point is that for normal processes, e.g. the case where
all moments of the waiting time PDF $\psi(\tau)$ exists, it does not matter how we start the process, 
in the long time limit the time and ensemble average procedures are all identical. 
In that sense the L\'evy walk, in the enhanced regime, is unique. 
A similar effect might be found also for sub-diffusive CTRW, for the case
of finite average waiting time, but with an infinite variance, but that is left for future work.
Further time averaged drifts, when a bias is present, also exhibit interesting ergodic features 
and Einstein relations, as we discussed recently in \cite{ours}.

\acknowledgement
This work was supported by the Israel Science Foundation.

\appendix

\section{Correlation functions}

For the derivation of the velocity correlation function we follow \cite{GodrLuck01}.
Hence, for the time variables $t_1$, $\Delta$ with $t_2 - t_1 = \Delta$ and going to Laplace domain
with respect to $\Delta$ we have
\begin{eqnarray}
\tilde p_n(t_1,u_{\Delta}) &=& 
\left\lbrace  
\begin{array}{l l }
\tilde \psi_{f,t_1}(u_\Delta) \tilde \psi^{n-1} (u_\Delta)\frac{1-\tilde \psi(u_\Delta)}{u_\Delta}\; \hspace{.2cm}  n \geq 1\\
\frac{1-\tilde \psi_{f,t_1}(u_\Delta)}{u_\Delta}\; \hspace{.85cm}   n = 0
\end{array}
\right. \label{pnoft}
\end{eqnarray}
where $\psi_{f,t_1}$ is the forward recurrence time PDF, i.e. the PDF of the time it takes to encounter the next event
after a given time $t_1$ which in double-Laplace domain reads
\begin{equation}
\tilde{\tilde \psi}_{f,{u_1}}(u_\Delta) = 
\frac{1}{1-\tilde{\psi}(u_1)} \frac{\tilde\psi(u_\Delta) -\tilde\psi(u_1)}{u_1 - u_\Delta} .\label{fw}
\end{equation}
Here $u_\Delta$ and $u_1$ are the Laplace variables conjugate to $\Delta$ and $t_1$, respectively. 
Inserting (\ref{pnoft}) and (\ref{fw}) back into (\ref{genCvv}) we get
\begin{eqnarray}
&&{\cal L}_{\Delta} \left\lbrace\langle v(t_1)v(t_1+\Delta)\rangle|u_\Delta\right\rbrace =
v_0^2 \frac{1-\tilde \psi_{f,t_1}(u_\Delta)}{u_\Delta} \nonumber\\
&& + v_0^2 \sum_{n=1}^\infty (-1)^n \tilde \psi_{f,t_1}(t_1,u_\Delta) \tilde \psi^{n-1}(u_\Delta) 
\frac{1 - \tilde{\psi}(u_\Delta)}{u_\Delta} \nonumber\\
&&= v_0^2 \left[ 
\frac{1-\tilde \psi_{f,t_1}(,u_\Delta)}{u_\Delta} - 
\tilde \psi_{f,t_1}(u_\Delta)\frac{1-\tilde\psi(u_\Delta)}{u_\Delta \left(1 + \tilde\psi (u_\Delta)\right)}
\right] , \nonumber \\
&&{\cal L}_{t_1,\Delta} \left\lbrace\langle v(t_1)v(t_1+\Delta)\rangle|u_1,u_\Delta\right\rbrace = \nonumber\\
&&
v_0^2 \left[ 
\frac{1}{u_\Delta u_1} - \frac{1}{1-\tilde\psi(u_1)}\frac{\tilde\psi(u_\Delta) - \tilde\psi(u_1)}{(u_1 - u_\Delta)u_\Delta}
\right.  \nonumber\\
&& \hspace{.2cm}
\left. -\frac{1}{1-\tilde\psi(u_1)}\frac{\tilde \psi (u_\Delta) - \tilde \psi (u_1)}{u_1 - u_\Delta}
\frac{1-\tilde \psi (u_\Delta)}{u_\Delta \left(1 + \tilde\psi (u_\Delta)\right)}
 \right] \nonumber \\
&=&  v_0^2 \frac{1}{u_\Delta}
\left[ \frac{1}{u_1}- \tilde{\tilde{\psi}}_{f,{u_1}}(u_\Delta)\frac{2}{1+\tilde{\psi}(u_\Delta)} \right] \label{CvvLap}
\end{eqnarray}
where we denote the Laplace transformation by ${\cal L}_{}$

\subsection{Ballistic phase}

Let us now specify the sojourn time distribution at large times, in Laplace domain 
$\tilde\psi(u) \simeq 1 - Au^\alpha $, $0<1<\alpha$ which yields
\begin{eqnarray}
&&{\cal L}_{t_1,\Delta} \left\lbrace\langle v(t_1)v(t_1+\Delta)\rangle|u_1,u_\Delta\right\rbrace =\nonumber\\
&&v_0^2 \left[ \frac{1}{u_{\Delta}u_{1}} - \frac{1}{1-\tilde{\psi}(u_1)} \frac{\tilde\psi(u_\Delta) -\tilde\psi(u_1)}{(u_1 - u_\Delta)u_{\Delta}}\right] \nonumber\\
&&=v_0^2 \left[\frac{1}{u_{\Delta}u_{1}} -\frac{1}{u_{\Delta}}\tilde{\tilde {\psi}}_{f,u_1}(u_{\Delta})\right]
\end{eqnarray}
and after Laplace inversion 
\begin{eqnarray}
\langle v(t_1)v(t_1+\Delta)\rangle  &=& v_0^2 \left( 1- \int_0^{\Delta} \psi_{f,t_1}(\Delta^\prime)\, d\Delta^\prime \right) \nonumber\\
&=& v_0^2 \int_\Delta^\infty \psi_{f,t_1}(\Delta^\prime)\, d\Delta^\prime ,
\end{eqnarray}
which we will use in the following since we know the scaling form of $\psi_{f,t_1}(\Delta)$ 
for large $t_1$ and $\Delta$ due to 
Dynkin's theorem Eq. (\ref{Dynkin}), leading to (\ref{Cvv}).

Inserting Eq. (\ref{Cvv}) into Eq. (\ref{Cxx0}) and using the definition of the incomplete Beta function
$B(y;a,b)=\int_0^y du {u^{a-1}(1-u)^{b-1}}$
and repeated integration by parts we find
\begin{eqnarray}
&&\langle x(t_1) x(t_2) \rangle = v_0^2 \frac{\sin \pi\alpha}{\pi} \times \nonumber\\
&&\Bigg[  
\int_0^{t_1} ds_1 \int_{s_1}^{t_2} B\left(\frac{t_1}{t_2};\alpha,1-\alpha \right)\,ds_2 \nonumber\\
&& + \int_0^{t_1} ds_2 \int_0^{s_2} B\left(\frac{t_1}{t_2};\alpha,1-\alpha \right) \, ds_1
\Bigg] \nonumber \\
&=&
v_0^2 \frac{\sin \pi\alpha}{\pi} \times 
\Bigg[ 
\int_0^{t_1} ds_1 \Big[ t_2 B\left(\frac{s_1}{t_2},\alpha,1-\alpha\right) \nonumber\\
&&- s_1 B\left(1,\alpha,1-\alpha\right) 
- s_1 B\left(\frac{s_1}{t_2},-1+\alpha,1-\alpha\right) \nonumber\\
&&+ s_1 B\left(1,-1+\alpha,1-\alpha\right) \Big]  \nonumber\\
&&
+\int_0^{t_1} ds_2 \Big[s_2(1-\alpha)\frac{\pi}{\sin\pi\alpha}\Big]
\Bigg]\nonumber \\
&=&
v_0^2 \frac{\sin \pi\alpha}{\pi} \times \nonumber\\
&&\Big[  
t_1 t_2 B\left(\frac{t_1}{t_2},\alpha,1-\alpha\right) - t_2^2 B\left(\frac{t_1}{t_2},1+\alpha,1-\alpha\right)\nonumber\\
&&
-\frac{1}{2}t_1^2 B\left(1,\alpha,1-\alpha\right) - \frac{1}{2}t_1^2 B\left(\frac{t_1}{t_2},-1+\alpha,1-\alpha\right) \nonumber\\
&&
+\frac{1}{2}t_2^2 B\left(\frac{t_1}{t_2},1+\alpha,1-\alpha\right) + \frac{1}{2}t_1^2 B\left(1,-1+\alpha,1-\alpha\right)
\Big]\nonumber\\
&&+\frac{v_0^2}{2}(1-\alpha) t_1^2
\label{cxxcalc}
\end{eqnarray}
which with $B(1,-1+\alpha,1-\alpha)=0$ for $0<\alpha<1$ finally yields Eq. (\ref{Cxx}).

\subsection{Subballistic phase}

Here the Laplace transform of the sojourn time of the particle in a velocity state for small $u$ reads
$\tilde{\psi(u)}=1-\langle\tau\rangle u + Au^{\alpha}$ with $\langle\tau\rangle=\alpha/{\alpha-1}$ and $A=|\Gamma(1-\alpha)|$.
For $t_1, \Delta$ large, i.e. $u_1,u_{\Delta}\to 0$ in Laplace domain, Eq. (\ref{CvvLap}) becomes again
\begin{eqnarray}
&&{\cal L}_{t_1,\Delta} \left\lbrace\langle v(t_1)v(t_1+\Delta)\rangle|u_1,u_\Delta\right\rbrace =\nonumber\\
&&v_0^2 \left[ \frac{1}{u_{\Delta}u_{1}} - \frac{1}{1-\tilde{\psi}(u_1)} \frac{\tilde\psi(u_\Delta) -\tilde\psi(u_1)}{(u_1 - u_\Delta)u_{\Delta}}\right] \nonumber\\
&&=v_0^2 \left[\frac{1}{u_{\Delta}u_{1}} -\frac{1}{u_{\Delta}}\tilde{\tilde {\psi}}_{f,u_1}(u_{\Delta})\right]\nonumber\\
&&
=v_0^2 \tilde{\tilde{p}}_0(u_1,u_{\Delta})
=v_0^2 \frac{A}{\langle\tau\rangle}\frac{u_1^{\alpha-1} - u_{\Delta}^{\alpha-1}}{u_1(u_1 - u_{\Delta})}.
\end{eqnarray}
Laplace inversion yields
\begin{eqnarray}
&&\langle v(t_1)v(t_1+\Delta)\rangle =\nonumber\\
&&=
v_0^2 \frac{A}{\langle\tau\rangle}
\frac{1}{(\alpha-1)|\Gamma(1-\alpha)|}
\left[
\Delta^{1-\alpha}-(t_1+\Delta)^{1-\alpha}
\right]\nonumber\\
&&=
v_0^2 \frac{A}{\langle\tau\rangle}
\frac{1}{(\alpha-1)|\Gamma(1-\alpha)|}
\left[
(t_2-t_1)^{1-\alpha}-t_2^{1-\alpha}
\right].
\end{eqnarray}
Inserting this again into Eq. (\ref{Cxx0}) gives
\begin{eqnarray}
&&\langle x(t_1)x(t_2)\rangle =\nonumber\\
&&=
v_0^2 \frac{A}{\langle\tau\rangle}
\frac{1}{\Gamma(4-\alpha)}
\left[
-(t_2-t_1)^{3-\alpha} + t_2^{3-\alpha} - (3-\alpha)t_2^{2-\alpha}t_1 + \alpha t_1^{3-\alpha}
\right],
\end{eqnarray}
which finally yields Eq. (\ref{CxxGL}).

\section{Ensemble averaged TAMSD, ballistic phase}

Inserting Eq. (\ref{Cxx}) and Eq. (\ref{msd}) into Eq. (\ref{TAMSD2}) and again using integration by parts 
and the integral definition of the incomplete Beta function yields
\begin{eqnarray}
&&\langle \overline{\delta^2}\rangle  = v_0^2\Bigg[\frac{1}{t-\Delta}\left[ \frac{1-\alpha}{3}(t^3 - \Delta^3)\right] \nonumber\\
&&- 
\frac{\sin \pi\alpha}{\pi (t-\Delta)}
\left[ \left(\frac 1 3 t^3 - \frac 1 2 \Delta t^2\right) B\left(\frac{t-\Delta}{t};\alpha,1-\alpha \right) \right. \nonumber\\
&&- 
\frac 1 3 \Delta^3B\left(\frac{t-\Delta}{t};\alpha, -2-\alpha\right) 
+ \frac 1 2 \Delta^3 B\left(\frac{t-\Delta}{t};\alpha,-1-\alpha \right) \nonumber\\
&&- 
\frac 1 6 t^3 B\left(\frac{t-\Delta}{t};1+\alpha,1-\alpha\right) \nonumber\\
&&+ 
\frac 1 6 \Delta^3 B\left(\frac{t-\Delta}{t};1+\alpha,-2-\alpha\right) \nonumber \\
&& + 
\left(-\frac 1 6 t^3 + \frac 1 2 \Delta t^2 - \frac 1 2 \Delta^2 t\right) 
B\left(\frac{t-\Delta}{t};-1+\alpha,1-\alpha\right) \nonumber\\
&&+ 
\frac 1 6 \Delta^3  B\left(\frac{t-\Delta}{t};-1+\alpha,-2-\alpha\right) \nonumber \\
&&
- \frac 1 2 \Delta^3 B\left(\frac{t-\Delta}{t};-1+\alpha,-1-\alpha\right) \nonumber \\
&&+ 
\left.\frac 1 2 \Delta^3 B\left(\frac{t-\Delta}{t};-1+\alpha,-\alpha\right)
\right]\Bigg],
\end{eqnarray}
For the small $\Delta$ expansion Eq. (\ref{EAtamsd}) we used the identity
\begin{equation}
B\left(\frac{t_1}{t_2},a,b \right) = B\left(1,b,a \right) - B\left( \frac{\Delta}{t_2},b,a \right)
\end{equation}
where $\Delta=t_2-t_1$
and the expansion of the incomplete Beta function for small arguments $y$
\begin{equation}
B\left( y,a,b \right) = y^a \sum_{n=0}^\infty \frac{\left(1-b \right)_n }{n!(a+n)}
\end{equation}
where $\left(c\right)_n = \Gamma(c+n)/\Gamma(c)$ is the Pochhammer symbol.



\end{document}